\input harvmac
%\draftmode
\def \lc {light-cone\ }

\def \const {{\rm const}}

 \def \four{{\textstyle{1\ov 4}}}
\def \ov {\over}

\def \ep {\epsilon}
\def \k {\kappa}

\def \ss {{\cal S}}

\def \pa { \partial}
\def \a {\alpha}
\def \E {{\cal E}}
\def \b {\beta}
\def \g {\gamma}
\def \G {\Gamma}
\def \d {\delta}

\def \l {\lambda}

\def \m {\mu}
\def \n {\nu}

\def \s {\sigma}

\def \ee {\epsilon}

\def \r {\rho}
\def \t {\theta}
\def \ta {\tau}
\def \p {\phi}

\def \vp {\varphi}

\def \ps {\psi}

\def \frac#1#2{{ #1 \over #2}}
\def \lr { \lref}
\def \td {\tilde}

\def \M {{\cal M}}
\def \aa {{\a'}}

\def \lr{\lref}

\def \M {{\cal M}}

\def \rf {\refs}

\def \ee {{\rm e}}

\def \adss {$AdS_5 \times S^5\ $}
\def \ads {$AdS_5$\ }

\def \pw {plane wave\ }

\def \lc {light-cone\ }
\def \ta { \tau}
\def \s { \sigma }

\def  \sqf {\l^{1/4}}

\def \vp {\varphi}

\def \p {\phi}
\def \vt {\theta}
 \def \a { \alpha}
\def \r {\rho}
\def \fourth {{1 \ov 4}}
\def \fo  {{{\textstyle {1 \ov 4}}}}

\def \inv {^{-1}}
\def \ri {{i}}

\def \vr {\varrho}
  \def \td { \tilde }

\def \t {\theta}

\def \del{\partial}
\def \m {\mu }
\def \n {\nu }
\def \ha { { 1 \over 2}}

\def \vr  { \varrho}

\def \g {\gamma}
\def \G {\Gamma}
\def \k {\kappa}
\def \l {\lambda}

\def \td {\tilde }
\def \b{\beta}

\def \ha {{1 \over 2}}
\def \ep{\epsilon}

\def \ov {\over}
 \def \Om {\Omega}

\def \H {{\cal H}}
\def \DE {\Delta E}

\def \w {w}
\def \sss {{\textstyle{ 2 \ss \ov \sqrt{1 + \nu^2}}}}

\def \om {\kappa}
\def \w  {\omega}
\def \oo  {\omega}
\def \ww { \nu }
\def \Ev {E_{\rm vac}}

\def \sql {{\sqrt{\l}}\ }
\def \hal {{\textstyle {1\ov 2}}}
\def \tri {{\textstyle {1\ov 3}}}

\def \ta {\tau}

\def \ap {\approx}

\def \isql {{\textstyle { 1 \ov \sqrt{\l}}}}
\def \rr {{\bar \r}}
\def \D {{\rm D}}
\def \E {{\cal E}}

\def \kk { {\textstyle { 1 \ov \l^{1/4}}}}

\lr \veg { H.~J.~de Vega and I.~L.~Egusquiza,
``Planetoid String Solutions in 3 + 1 Axisymmetric Spacetimes,''
Phys.\ Rev.\ D {\bf 54}, 7513 (1996)
[hep-th/9607056].
%%CITATION = HEP-TH 9607056;%%
 }

 \lr \mah { S.~Kar and S.~Mahapatra,
``Planetoid strings: Solutions and perturbations,''
Class.\ Quant.\ Grav.\  {\bf 15}, 1421 (1998)
[hep-th/9701173].
%%CITATION = HEP-TH 9701173;%%
 }

  \lr \ves { H.~J.~de Vega, A.~L.~Larsen and N.~Sanchez,
``Semiclassical quantization of circular strings in de Sitter and
anti-de Sitter space-times,''
Phys.\ Rev.\ D {\bf 51}, 6917 (1995)
[hep-th/9410219].
%%CITATION = HEP-TH 9410219;%%
}

\lr \oog {  G.~T.~Horowitz and H.~Ooguri,
``Spectrum of large N gauge theory from supergravity,''
Phys.\ Rev.\ Lett.\  {\bf 80}, 4116 (1998)
[hep-th/9802116].
%%CITATION = HEP-TH 9802116;%%
E.~Witten,
``Anti-de Sitter space and holography,''
Adv.\ Theor.\ Math.\ Phys.\  {\bf 2}, 253 (1998)
[arXiv:hep-th/9802150].
%%CITATION = HEP-TH 9802150;%%
T.~Banks, M.~R.~Douglas, G.~T.~Horowitz and
E.~J.~Martinec,
``AdS dynamics from conformal field theory,''
hep-th/9808016.
%%CITATION = HEP-TH 9808016;%%

}

\lr\fro{
A.~L.~Larsen and V.~P.~Frolov,
``Propagation of perturbations along strings,''
Nucl.\ Phys.\ B {\bf 414}, 129 (1994)
[hep-th/9303001].
%%CITATION = HEP-TH 9303001;%%
}

\lr \bmn { D.~Berenstein, J.~Maldacena and H.~Nastase,
``Strings in flat space and pp waves from N = 4 super Yang Mills,''
JHEP {\bf 0204}, 013 (2002)
[hep-th/0202021].
%%CITATION = HEP-TH 0202021;%%
}

\lr \gkp { S.~S.~Gubser, I.~R.~Klebanov and A.~M.~Polyakov,
``A semi-classical limit of the gauge/string correspondence,''
hep-th/0204051.
%%CITATION = HEP-TH 0204051;%%
}

\lr \pol {A.~M.~Polyakov,
``Gauge fields and space-time,''
hep-th/0110196.
%%CITATION = HEP-TH 0110196;%%
}

\lr \blau{M.~Blau, J.~Figueroa-O'Farrill, C.~Hull and G.~Papadopoulos,
``A new maximally supersymmetric background of IIB superstring theory,''
JHEP {\bf 0201}, 047 (2002)
[hep-th/0110242].
%%CITATION = HEP-TH 0110242;%%
``Penrose limits and maximal supersymmetry,''
hep-th/0201081.
%%CITATION = HEP-TH 0201081;%%
}

\lr \marey {J.~Maldacena,
``Wilson loops in large N field theories,''
Phys.\ Rev.\ Lett.\  {\bf 80}, 4859 (1998)
[hep-th/9803002].
%%CITATION = HEP-TH 9803002;%%
S.~J.~Rey and J.~Yee,
``Macroscopic strings as heavy quarks in large N gauge theory and
anti-de Sitter supergravity,''
Eur.\ Phys.\ J.\ C {\bf 22}, 379 (2001)
[hep-th/9803001].
%%CITATION = HEP-TH 9803001;%%
}

\lref\mets{
R.~R.~Metsaev and A.~A.~Tseytlin,
``Exactly solvable model of superstring in plane wave Ramond-Ramond
background,''
hep-th/0202109.
%%CITATION = HEP-TH 0202109;%%
}

\lref\mett{
R.~R.~Metsaev,
``Type IIB Green-Schwarz superstring in plane wave Ramond-Ramond
background,''
Nucl.\ Phys.\ B {\bf 625}, 70 (2002)
[hep-th/0112044].
%%CITATION = HEP-TH 0112044;%%
}

\lref\mt{
R.~R.~Metsaev and A.~A.~Tseytlin,
``Type IIB superstring action in \adss  background,''
Nucl.\ Phys.\ B {\bf 533}, 109 (1998)
[hep-th/9805028].
%%CITATION = HEP-TH 9805028;%%
}

\lr \dgt {N.~Drukker, D.~J.~Gross and A.~A.~Tseytlin,
``Green-Schwarz string in \adss : Semiclassical partition
function,''
JHEP {\bf 0004}, 021 (2000)
[hep-th/0001204].
%%CITATION = HEP-TH 0001204;%%
A.~A.~Tseytlin,
``'Long' quantum superstrings in \adss ,''
hep-th/0008107.
%%CITATION = HEP-TH 0008107;%%
}
\lref\kal{
R.~Kallosh and A.~A.~Tseytlin,
``Simplifying superstring action on \adss,''
JHEP {\bf 9810}, 016 (1998)
[hep-th/9808088].
%%CITATION = HEP-TH 9808088;%%
}

\lr \kall {R.~Kallosh, J.~Rahmfeld and A.~Rajaraman,
``Near horizon superspace,''
JHEP {\bf 9809}, 002 (1998)
[hep-th/9805217].
%%CITATION = HEP-TH 9805217;%%
R.~Kallosh and J.~Rahmfeld,
``The GS string action on \adss ,''
Phys.\ Lett.\ B {\bf 443}, 143 (1998)
[hep-th/9808038].
%%CITATION = HEP-TH 9808038;%%
}

\lr \theis  { S.~Forste, D.~Ghoshal and S.~Theisen,
``Stringy corrections to the Wilson loop in N = 4 super Yang-Mills
theory,''
JHEP {\bf 9908}, 013 (1999)
[hep-th/9903042].
%%CITATION = HEP-TH 9903042;%%
``Wilson loop via AdS/CFT duality,''
hep-th/0003068.
%%CITATION = HEP-TH 0003068;%%
}

\lref\bars{
I.~Bars,
``Folded strings in curved space-time,''
hep-th/9411078.
%%CITATION = HEP-TH 9411078;%%
}

%%%%%%%%%%%%%%%%%%%%%%%%%%%%%%%%%%%%%%%%%%%%%%%%%%%%%
\Title{\vbox
{\baselineskip 10pt
{\hbox{         }
}}}
{\vbox{\vskip -30 true pt
\centerline { Semiclassical quantization  }
\medskip
\centerline {of rotating  superstring in $AdS_5 \times S^5$
 }
\medskip
\vskip4pt }}
\vskip -20 true pt
\centerline{S. Frolov$^{a,}$\footnote{$^*$} {Also at Steklov
Mathematical Institute, Moscow.}
 and
A.A.~Tseytlin$^{a,b,}$\footnote{$^{**}$}
{Also at
Lebedev Physics Institute, Moscow.}
}
\smallskip\smallskip
\centerline{ $^a$ \it  Department of Physics,
 The Ohio State University,
 Columbus, OH 43210, USA}

\centerline{ $^b$ \it  Blackett Laboratory,
 Imperial College,
 London,  SW7 2BZ, U.K.}

\bigskip\bigskip
\centerline {\bf Abstract}
\baselineskip12pt
\noindent
\medskip

Motivated by recent  proposals in   hep-th/0202021 and
hep-th/0204051  we develop  semiclassical
quantization of superstring  in $AdS_5 \times S^5$. We
start with  a  classical solution describing
 string rotating in  $AdS_5$  and boosted  along
 large circle  of $S^5$. The energy
 of the classical solution $E$  is a  function  of
 the spin $S$ and  the momentum  $J$ (R-charge)
 which interpolates  between the limiting cases
  $S=0$ and $J=0$ considered
 previously. We  derive the  corresponding
 quadratic fluctuation  action  for bosonic and fermionic
 fields from the  GS string action  and
  compute the string  1-loop (large $\sqrt \lambda= {R^2\ov  \a'}$)
   correction to the classical energy
  spectrum in the $(S,J)$ sector.
 We find that the 1-loop  correction to the
 ground-state energy
 does not cancel for $S\not=0$.
 For large $S$  it scales as  $\ln S$, i.e.
 as  the classical  term,
 with no higher powers of $\ln S$ appearing. This
   supports  the conjecture made in  hep-th/0204051
 that the classical $E-S = a \ln S$ scaling can
  be interpolated  to weak coupling  to  reproduce the
  corresponding  operator
 anomalous dimension behaviour in gauge theory.

\bigskip

%%%%%%%%%%%%%%%%%%%%%%%%%%%%%%%%%%%%%%%%%%%%%%%%%%%%%%%%%
\Date{04/02}

%%%%%%%%%%%%%%%%%%%%%%%%%%%%%%%%%%%%%%%%%%%%%%%%%%%%%%%%%%%%%%%%%%%
\noblackbox
\baselineskip 16pt plus 2pt minus 2pt
%\baselineskip 20pt plus 2pt minus 2pt

%%%%%%%%%%%%%%%%%%%%%%%%%%%%%%%%%%%
\newsec{Introduction}
%%%%%%%%%%%%%%%%%%%
Superstring  theory in \adss is a  highly symmetric
yet complicated-looking theory: both bosonic and fermionic
sectors of the  corresponding 2-d action
are  non-linear  and   hard to quantize directly.
This precludes one from  proving  the
expected  duality to  $\cal N$=4 SYM theory
(even  in the large $N$ limit)
in  a  direct way.
It was recently suggested \refs{\pol,\bmn,\gkp}
that looking at special sectors of   states
in the full string spectrum  where
 certain  quantum  numbers  are  large
may allow one to  check the AdS/CFT correspondence
beyond the supergravity level.

In the simplest  and most explicit  example
  \bmn\  one considers the
sector of string states  localized  at  the center of
$AdS_5$ and  carrying large momentum $J$
along the central circle  of  $S^5$.
The leading (1-loop) term in the energy  spectrum
of the corresponding  {\it quantum} string oscillations
can be found  explicitly
in the limit  of large  string tension
or $\sql = { R^2\ov \a'} \gg 1$.
Using the  near-BPS nature of these string states
one  is  then  able to  reproduce \bmn\  the same
string  oscillator spectrum
on the gauge-theory side  as the spectrum of
anomalous dimensions of single-trace  operators carrying large
R-charge.

In a more complicated but  potentially more  interesting
case  \gkp\ one concentrates on  states
with  large angular momentum  $S$ in  \ads
(corresponding  to spin  in the boundary theory).
A remarkable observation \gkp\
is that the {\it classical}   energy  of a rotating  string
in \ads space  scales,
 for large ${S\ov \sql} \gg 1 $,
 not as in flat  (Regge) case  $E= \sqrt{ 2\l S} $ but
as $E= S + {\sql\ov \pi} \ln {S\ov \sql}  + ...$,
 which looks the same  as the $S + a \ln S$ behaviour
of the  (canonical+anomalous)  dimension of the
corresponding
bilinear  operator in gauge-theory
perturbative expansion.
Since the  string rotating in \ads is not a BPS state,
one  expects that its energy should receive
quantum   (sigma-model loop) corrections,
\eqn\ink{
E= S  +  f(\l) \ln { S\ov \sql}   +
 h(\l) \ln^2  { S\ov \sql}   + ...  \ , }
where  $f,h,...$ are  given by series
in $\l \gg 1$ (large tension) expansion
\eqn\nnn{
f(\l)  =  {\sql\ov \pi}  + a_1  +
{ a_2 \ov \sql } + O({ 1  \ov \l })   \ , \ \ \ \ \ \ \
h(\l) =  k_1  +
{ k_2 \ov \sql } +   O({ 1  \ov \l  }) \ , ... \  \ . }
To be able to establish a  correspondence
 with gauge-theory results
 $f(\l)$ should admit an interpolation to
weak coupling,
  $f(\l)_{\l \ll 1} = b_1 \l + b_2
\l^2 + ...$, which  should agree  with the  SYM perturbation
theory.    For consistency
of the  proposal of \gkp\  all  stronger that $\ln S$, i.e.
 $\ln^m S$ ($m=2,3,...$),
corrections  should cancel, both on the   gauge-theory
\gkp\   and the string-theory  side.

Here we shall compute the 1-loop  string  correction
to the energy of the  classical rotating \ads
string solution \veg\ considered  in \gkp\
and show that while $f(\l)$ in \nnn\
does receive  a correction ($a_1$ is non-zero),
$h(\l)$ indeed vanishes  ($k_1=0$).
%NN
We shall also argue that the same remains to be true
to all orders in string $1\ov \sqrt \l$ expansion.
This provides strong support to the suggestion of \gkp,
and thus  another highly non-trivial check of the
string theory -- gauge theory correspondence.

Below we shall develop the general formalism
for  semiclassical  quantisation
near classical sigma-model
solution describing rotating string
 in \adss, with the aim to compute
 the leading (1-loop)
 term in the energy of  excited quantum string states
 in the sector with  given  angular momenta.
We  will  use the  Green-Schwarz  formulation \mt,  and
to the 1-loop  order,   will need
only the  simple  quadratic
term   in  the full fermionic action.
Part of the discussion will be  very similar
to  the one in \refs{\kal,\theis,\dgt} where the  1-loop
correction to the \adss string partition function
near a ``long'' string  configuration relevant
for the Wilson loop average \marey\ was computed.

We shall  consider the classical rotating  solution  that
generalises  both special cases considered
in \gkp\ and \bmn, i.e. describes a   folded  closed  string
stretched  along radial direction $\r$ of \ads and
rotating  along  large  circle $\p$ of the boundary $S^3$,
 with its center of mass
 moving   along the  large
circle  $\vp$ of $S^5$.  The  classical  energy
is a complicated function  of the
two angular momenta $S$ and $J$.
 It  interpolates  between
$E= J$ for $S=0$,  $E\ap S + {\sql \ov \pi}   \ln S $  for $J=0$
and  large $S$, and
$E\ap  \sqrt {J^2 +  2\sqrt { \l} S}$   for  ${J\ov \sql}\ll 1,\
{S\ov \sql}  \ll 1$
(which is   the  familiar   flat-space  expression
$E^2 = J^2 + {2\ov \a'} S $).
 Consideration of
this  2-parameter case is illuminating as it allows
to  interpolate  between
   part of the quantum string spectrum
in \bmn\ (with spin represented  by small
string oscillations
in \ads directions)  and  the classical  spectrum
of a short string rotating in \ads and boosted along
the circle of $S^5$  (see section 3).

We shall start the discussion  of the 1-loop
quantum  correction
to the classical energy spectrum in section 4
with the case  of $J\not=0$ but  $S=0$, i.e. with
non-zero $S^5$ boost but  no spin in \ads. We shall
explain  in detail (following suggestions  of
\bmn\ and  \gkp)
that  the semiclassical quantization  near  the boosted
point-like string  solution   gives directly the
 same oscillator string spectrum as
 obtained by  exact quantization  \refs{\mett,\mets}
 of the closed  superstring
 in the plane-wave \blau\ background.
Since in the 1-loop approximation  near a bosonic
string solution
one  uses only the quadratic term in the fermionic
part of the action, one  does not need to go through the
derivation
 \mett\ of the exact form of the GS action  in the \pw
 background, and the choice of the
 \lc kappa-symmetry gauge is essentially imposed on us
by the  form of the classical string  background.

In section 5  we shall  derive the action for small
fluctuations near the general (boosted and rotated)
string solution, by considering bosonic fields  in both
static (sect. 5.1)
and conformal (sect. 5.2)  gauges and
obtaining (in sect. 5.3) the fermionic part
of the action  from the quadratic part of the
GS action in \adss background
(for non-zero $S$ we can  fix covariant
kappa-symmetry gauge as in \dgt).
The resulting action describes a collection of 2-d
bosonic and fermionic fields  on a flat cylinder
$(\tau, \sigma)=(\tau,\sigma+ 2\pi)$   having, in general,
  non-constant
($\s$-dependent) masses.

In section 6 we shall  study   the 1-loop
quantum corrections to the energy
 of the classical solution  using certain approximations.
 We shall first compute the 1-loop
 shift  of the  ground-state energy $\Delta E$ (sect. 6).
 It   cancelled  out  in the BPS case $E=J, \ S=0$ \bmn\
 but  does not  vanish  in the case of
 non-zero spin. In the limit of large $S$ and $J=0$ we shall
  find
  that the 1-loop coefficients  in \nnn\ are
 $a_1 \ap  - { 3 \ln 2 \ov  \pi}  $ and $k_1 =0$.
%NN
We shall also argue that  in general
string $\alpha'$  corrections will never produce   higher
than $\log S$ corrections to the energy.

We shall then  consider  (in sect. 6.2) the excited
oscillator   part of the 1-loop  string  spectrum
(i.e. a generalization of the oscillator  spectrum
in the $S=0$ case \bmn)
 by
considering  special  limits in the $(S,J)$ parameter
space.
It would be  interesting to see  if the
anomalous  dimensions of gauge-theory operators with large spin
and large  R-charge  have similar   scaling
with $S$ and $J$ as  we find
 on  the string-theory side.

% The form of the resulting spectrum
 %suggests, in particular,  that the  $\ln S$ term
 %\gkp\  in the anomalous
% dimensions of the
%  gauge-theory operators  with
% large R-charge should be suppressed.

Appendix A  contains a  derivation
of the general
expression   for the correction to the
 energy  of  classical rotating string solution
in terms of the 2-d  Hamiltonian for quadratic  fluctuation
fields using the conformal-gauge constraints.

%%%%%%%%%%%%%%%%%%%%%%%%%%%%%%%%%%%%%%%%%%%%%%%%%%%%%%%%
%%%%%%%%%%%%%%%%%%%%%%%%%%%%%%%%%%%%%%%%%%%%%%%%%%%%%%%%

%%%%%%%%%%%%%%%%%%%%%%
%%%%%%%%%%%%%%%%%%%%%%%%%%%%%%%%%%%%%%%%%%%%%%%%%%%%%%%%
\newsec{Superstring action in \adss }
%%%%%%%%%%%%%%%%%%%%%%%%%%%%%%%%%%%%%

Our starting point  will  be the GS superstring action
in \adss written in the following general form
\eqn\gss{
I= - { \sql  \ov 2 \pi }
\int d^2 \xi  \  \big[ L_B (x,y)
+ L_F ( x, y, \theta)  \big] \ , \ \ \ \ \ \ \ \
\sql \equiv  { R^2 \ov  \aa}}
\eqn\kok{
L_B=
 \ha \sqrt { -g} g^{ab}   \big[
G^{(AdS_5)}_{mn}(x) \del_a x^m \del_b  x^n
+ G^{(S^5)}_{m'n'}(y) \del_a y^{m'} \del_b y^{n'} \big] \ . }
Here  $\xi^a=(\tau,\s), \ \s\equiv \s + 2 \pi$.
  We shall use Minkowski signature  in both
target space and  world sheet,  so that in
conformal gauge
$\sqrt {-g} g^{ab} = \eta^{ab}=$diag(-1,1).
The fermionic part is
\eqn\fer{
L_F=
i (\sqrt {- g} g^{ab }\delta^{IJ} -
\ep^{ab } s^{IJ} ) \bar \vt^I \vr_a D_b \vt^J   + O(\theta^4)\ ,
}
where
 $I,J=1,2$,  $s^{IJ}=$diag(1,-1),
 $\vr_a$ are projections of the 10-d Dirac
matrices,
\eqn\hop{
\vr_a \equiv \G_{A} E^{A}_M \del_a X^M = ( \G_p
E^p_M
+\G_{p'}  E^{p'} _M ) \del_a X^M \ , }
 and $E^{A}_M$ is
the
vielbein of the 10-d target space metric
(see \refs{\mt,\dgt} for  details).
Here $X^M= (x^m,y^{m'})$ are the  coordinates
and
$p,p'$ are the tangent space indices of \ads and $S^5$.
The covariant derivative $D_a$  is the projection
$\del_a X^M D_M $ of the
10-d derivative
$D^{IJ}_{ M}= (\del_{ M}
+ \fourth\omega^{AB}_{ M} \Gamma_{AB}) \delta^{IJ}
 - { 1 \ov 8 \cdot 5!}
  F_{ A_1... A_5} \G^{ A_1...A_5} \G_{ M}\epsilon^{IJ}  $.
Since $\theta^I$  are  10-d MW  spinors of the same
chirality   and since $F_5 = \ep_5 + * \ep_5$ for the
 \adss background, it
can be put into  the  following form
\eqn\form{
D_a\t^I   = (\delta^{IJ} {\D}_a
- { i \ov 2 } \epsilon^{IJ}  \G_* \vr_a ) \vt^J\ ,
\ \ \ \ \ \  \G_* \equiv i \G_{01234} \ , \ \
\G_*^2 =1 \ ,
 }  where
 $ {\D}_a = \del_a
+\fourth \del_a  X^M \omega^{AB}_M\Gamma_{AB
}$ and the ``mass term''  originates from the R-R coupling (cf.
\refs{\mt,\kal}).\foot{In the special ``5+5''  representation
of the $\G$-matrices used in \dgt\
one can write
$ D_a\t^I = (\delta^{IJ} {\D}_a
- {\ri\ov 2 } \epsilon^{IJ} \tilde\vr_a ) \vt^J, $
\  $ \tilde\vr_a\equiv \G_p E^p_M +\ri\G_{p'} E^{p'}_M $.}

In the leading large $\sql={ R^2 \ov  \aa} $
approximation near a classical bosonic
solution
we will be discussing we will
need  to know only the quadratic fermionic term.\foot{The
full form
of the GS string action in \adss
  was  explicitly written down \refs{\mt,\kall}
in  the Poincare coordinates of \ads.}
Our approach   will be  similar  to the one  in
\refs{\kal,\theis,\dgt}  where  1-loop string correction
to the Wilson loop factor (i.e. long open string
configuration)  \marey\  was discussed.
Here we will
follow \refs{\bmn,\gkp} and will
need to use global
coordinates in which the space-time energy of a string state
is  directly related to the dimension of the corresponding state
in dual gauge theory \oog.

We shall use the following explicit parametrization of
the (unit-radius)  metrics of
\ads and $S^5$:
\eqn\add{
(ds^2)_{AdS_5}
=  G^{(AdS_5)}_{mn}(x) dx^m  dx^n
= - \cosh^2 \r \ dt^2 +  d\r^2 + \sinh^2\r \ d\Om_3 \ , }
$$
d\Om_3
= d \b_1^2 + \cos^2 \b_1 ( \ d\b_2^2 +
 \cos^2 \b_2  \ d \beta_3^2)   \ , \ \ \ \ \   \beta_3\equiv \p \ . $$
\eqn\ade{
(ds^2)_{S^5}
=  G^{(S^5)}_{m'n'}(y) dy^{m'}  dy^{n'}
= d \ps_1^2 + \cos^2 \ps_1 ( \ d\ps_2^2 +
\cos^2 \ps_2\   d \Om_3')  \ , }
$$
d\Om'_3
= d \ps_3^2 + \cos^2 \ps_3 \ ( d\ps_4^2 +
 \cos^2 \ps_4 \  d \psi_5^2 )  \  , \ \ \ \ \  \   \ps_5\equiv \vp \ .
   $$

%%%%%%%%%%%%%%%%%%%%%%%%%%%%%%%%%%%%%%%%%%%%%%%%%%%%%%%%
\newsec{Classical solution for rotating string
in \adss }
%%%%%%%%%%%%%%%%%%%%%%%%%%%%%%%%%%

The classical string solution we shall consider
is a direct generalisation of
both the particle (point-like string)  moving  with  the speed
 of light along the $\vp$-circle of $S^5$
 \bmn\ and the  solution of \gkp\
describing a  closed string rotating  in the
$(\r,\p)$ plane of \ads
(this  solution was originally found in \veg,  and
was discussed
also in  \mah).\foot{Other
oscillating circular string solutions in AdS space
 were  studied    in \ves.}
 It is  straightforward to check that the following
rotating  and boosted closed string configuration solves
the classical string
equations in \adss  space
$$
t= \om \ta \ , \ \ \ \p= \w \tau \ , \ \ \
 \ \ \vp=   \ww \tau \ ,  \ \ \   \  \
 \om,\w,\ww =\const \ , $$
 \eqn\sll{ \r= \r (\s) = \r(\s + 2\pi) \ , \ \  \ \
\  \b_i=0 \ \ (i=1,2) \ , \ \ \  \ \ \
 \ps_s =0 \ \ (s=1,2,3,4) \ , }
where  $\r$ is subject to the corresponding  second-order
 equation (prime denotes  derivative over $\s$)
 \eqn\secc{
 \r''= (\k^2-\w^2) \sinh\r \ \cosh \r  \ . }
The   first of the   conformal gauge  constraints
\eqn\conn{
 G^{(AdS_5)}_{mn}(x) ( \del_0 x^m \del_0  x^n
+ \del_1 x^m \del_1  x^n)
+ G^{(S^5)}_{m'n'}(y) (\del_0 y^{m'} \del_0 y^{n'}
+  \del_1 y^{m'} \del_1 y^{n'} )  =0 \ , }
\eqn\coonn{
 G^{(AdS_5)}_{mn}(x) \del_0 x^m \del_1  x^n
+ G^{(S^5)}_{m'n'}(y) \del_0 y^{m'} \del_1 y^{n'}  =0 \ , }
then  says   that $\r(\s)$ must satisfy  the
following 1-st order equation (implying
\secc\ for $\r'\not=0$)\foot{In terms of the coordinate
$r=\tanh { \r\ov 2}$ this equation can be interpreted
as describing periodic motion in the central
part of the inverted double-well potential.}
\eqn\rgo{
\r'^2 =  \k^2 \cosh^2 \r - \w^2 \sinh^2 \r -  \ww^2  \
 \ . }
The periodicity  condition on  $\r(\s)$
is satisfied  by considering a folded string configuration.
In the simplest (``one-fold'') case considered in \gkp\ \foot{Multifold
configurations correspond to states with lower spin for given
energy.}
the interval $0\leq \s < 2\pi$ is split into 4 segments:
for $ 0< \s < \pi/2$ the function  $\r(\s)$
increases  from 0  to  its maximal value $ \r_0 $ \
($\r' (\pi/2)=0$)
\eqn\hmm{
(\k^2 -\ww^2) \cosh^2 \r_0  - (\w^2
- \ww^2)  \sinh^2 \r_0 = 0 \ ,  }
then  for $ \pi/2 < \s < \pi$
 decreases to zero, etc.\foot{In flat space
 $ds^2 = - dt^2 + d\r^2 + \r^2 d\p^2 + ... $  the
 corresponding rotating  string solution  (with $\nu=0$)
  is described  by  $ t= \k \tau, \ \p= \w \tau\ , \
  \r= \r_0 |\sin \w \s | $, $ \r_0= { \k\ov \w} $,
  so  that $ \r(\s) = \r(\s + 2 \pi) $ implies that
  $ \w$ must be  an integer (=1 in the one-fold case).
  The modulus accounts for the fact  that $\r$ must be positive.
  An  apparent singularity  at $\s=0$ (i.e. $\r=0$)
   is  simply a  coordinate one:  the solution
   has the standard  regular form when  written in the cartesian
   coordinates in the 2-plane:
   $x_1 \equiv \r \cos \p =  \r_0 \sin \w \s \cos \w \tau , \ \
   x_2 \equiv \r \sin \p =  \r_0 \sin \w \s \sin \w \tau $.}
The periodicity thus implies an extra
condition on the parameters:
\eqn\joi{
2 \pi = \int^{2\pi}_0 d \s
= 4 \int^{\r_0}_0   { d \r \ov
\sqrt{     (\k^2 -\ww^2) \cosh^2 \r
-
(\w^2 - \ww^2)  \sinh^2 \r    }   } \
. }
The corresponding induced 2-d metric on the $(\tau,\s)$
cylinder  is  conformally-flat
\eqn\coop{
g_{ab} = G_{MN} (X) \del_a X^M \del_b X^N
 = \r'^2(\s)  \eta_{ab} \ .}
Its   curvature
\eqn\cuu{
R^{(2)} = -  \r'^{-2} \del^2_\s \ln \r'^2
= -2  +   { 2 (\k^2-\n^2) ( \w^2 - \n^2) \ov \r'^4 } \  }
has  singularities on the   lines  $\s= \pi/2$ and
 $\s= 3\pi/2$, i.e. at  the turning points  where
 $\r'=0$.\foot{
 %NN
 The induced metric
 written in terms of $\r$ and $\tau$ as the coordinates
 is $ds^2 =  - f(\r) d\tau^2 +     d\r^2$ where
 $f=\r'^2 = \k^2 \cosh^2 \r - \w^2 \sinh^2 \r -  \ww^2 $.
 The turning points are thus  singularities coinciding with the
 horizons. For a
general  discussion of  folded classical string solutions
  in curved $D \geq  2$ space and related references
   see \bars.}

When $\ww=0$, i.e. when the rotation is  only in \ads  space,
this is the case considered in \refs{\gkp,\veg}.
If instead  we  choose  $\w=0$,
then  to satisfy  the periodicity
condition we must  set $\r=\r_0=\const$, but
then \secc\ implies that one should put
$\r_0=0$. \foot{A possible alternative is
not to impose periodicity in $\s$ and   consider
instead  an open string stretched all the way to the boundary of
\ads.}  Thus  we get a string shrunk
to a point, placed  at the
center  of \ads and   rotating
along the  circle in $S^5$  with the speed  of light
\rf{\bmn,\gkp}.

Another special case  is
 when $\k=\pm \w$: this  leads to $\r''=0$
 and $\r'^2 = \k^2 - \nu^2$=const.
 To have  a continuous periodic $\r(\s)$
  we need to demand $\k=\pm \nu$, i.e.  $\r=\r_0$=const.
  This case is
  equivalent to $\r=0$ and $\w=0, \ \k=\pm \nu$
  by a coordinate transformation in \ads
  (a combination of $SO(1,1)$ and $SO(2)$ rotation).

 Since the  solution depends only on the two combinations
$\k^2 -\ww^2$   and   $ \w^2 - \ww^2$,
the analysis of its properties
 is very similar to the one in  \gkp.
Introducing  the parameter $\eta$
\eqn\too{
 \coth^2 \r_0 = {\omega^2-\nu^2\ov \k^2-\nu^2}  \equiv
 1 + \eta \ ,\ \ \ \ \ \ \ \   \eta > 0 \ ,
 }
 it  is not difficult to see that solutions
  with finite energy exist
if (we  choose $\k,\w,\n$  to be positive)
\eqn\uio{\k >\nu\ ,\ \ \ \  \ \ \ \  \omega > \nu\ . }
 As in \gkp\ the  three conserved momenta  conjugate  to $t,\p,\vp$
 are the space-time energy $E$  and the two angular momenta
 $S$  and $J$:
 \eqn\eee{
 E= P_t= {\sql  } \om \int^{2\pi}_0  {d \s\ov  2 \pi} \ \cosh^2 \r
 \equiv \sql \E
\ ,   }
\eqn\spi{
S= P_\p= {\sql   } \w \int^{2\pi}_0 {d \s\ov  2 \pi}
 \ \sinh^2 \r \equiv \sql \ss
\ , }  \eqn\jje{
 J= P_\vp=   {\sql  } \ww
  \int^{2\pi}_0 {d \s\ov  2 \pi}  \  = \sql  \ww
  \ .   }
Note the   relation
\eqn\ES{
{\E} = \k  + {\k \ov\oo}  \ss    \ ,}
which  together with \joi\
  may  be used to determine the dependence of $E$
on $S$ and $J$.

In the full quantum theory $S$ and $J$ should take quantized
values.
In the semiclassical approximation we shall consider
we shall assume  that  their values are very large,
i.e. that  $  S\ov \sql $ and $J\ov \sql $ are finite for
$\sql \gg 1$, or, equivalently, that  the parameters
of the classical solution $\k,\w,\n$ are fixed  in the limit
of large  $\sql $.

Computing the integrals in \joi,\eee,\spi\
(using $d\s = \r'^{-1}  d\rho$
and \rgo)   we find
\eqn\jjei{
\sqrt{\k^2 -\ww^2} =  {1 \ov\sqrt{\eta}}\
{}_2F_1\left({1\ov 2},{1\ov 2};1;- {1\ov \eta}\right)
\ , }
 \eqn\ssei{
 \E  = {\om\ov \sqrt{\k^2 -\ww^2}} {1  \ov\sqrt{\eta}}\
{}_2F_1\left(-{1\ov 2},{1\ov 2};1;- {1\ov \eta}\right)
\ , \ \ \ \ \ \
\ss= {\oo\ov \sqrt{\k^2 -\ww^2}} {1 \ov 2 \eta \sqrt{\eta}}\
{}_2F_1  \left({1\ov 2},{3\ov 2};2;- {1\ov \eta}\right)
\ . }

Let us now  follow \gkp\ and consider
the limits of  ``short'' ($\r_0 \to 0$, i.e.
$\eta \to \infty$)   and ``long''
  ($\r_0 \to \infty$, i.e.
$\eta \to  0 $)  strings.

 %%%%%%%%%%%%%%%%%%%%%%%%%%%%%%%%%%%%%%%%%
 \subsec{\bf Short string}
%%%%%%%%%%%%%%%%%%%%%%%%%%%%%%%%%%%

For
 $\eta \gg 1$ we
 get from \jjei\ and \too
 \eqn\kapsh {
\k^2 \approx \nu^2 + {1\ov \eta} \ , \ \ \ \ \ \ \ \  \
 \oo^2 \approx \k^2 + 1 \approx
  \nu^2 + 1 +  {1\ov \eta}\ .
}
Taking into account that $\oo^2-\k^2\approx 1$,
an approximate solution  describing a short
rotating  string  has $\r(\s)$ satisfying
\eqn\shortsol{
\sinh^2\rho \approx \r^2  \approx {1\ov\eta}\sin^2\s\ .}
Eq.\ssei\ implies
\eqn\etash{
{1\ov \eta}\approx {2\ss\ov\sqrt{1+\nu^2}} \ll  1 \ ,
\ \ \ \ { \rm i.e.} \ \ \  \ \ \ 4 \ss^2 \ll  1+ \n^2   \ . }
Eq.\ES\ then determines  the dependence of  the energy
$E$ on the angular momenta  $J= \sql \nu $ and $S=\sql  \ss $
\eqn\Eshort{
\E \approx \sqrt{\nu^2   + {\sss}  }\  + \
 \sqrt{
{ \nu^2  + \sss
\ov 1+\nu^2 +
\sss}  }\ \  \ss \  .}
This relation is valid for any $\nu$
and $\ss$ satisfying \etash.
Further simplification  depends on a  particular
 value of   $\n$.
If $\nu$ is small, i.e. $\nu  \ll 1$  then $\ss \ll 1$
and \Eshort\ becomes
\eqn\hort{
\E \approx \sqrt{\nu^2   + 2 \ss  + ...   }\    + ...
\ ,\ \ {\rm  i.e.}  \ \ \ \ \ E^2\approx J^2 + 2\sqrt{\l} S + ...\ . }
This  expression has a simple interpretation.
The limit of short  strings  probes  small-curvature region of
\ads  where   $\r\approx 0$, i.e.  the second  term in the r.h.s.
of \ES\ and thus the second square root in \Eshort\
should be small, and the energy spectrum
should be approximately the same as in flat space.
Indeed, \hort\ is  the correct relativistic  expression
for the  energy of a string in flat space
 moving   along a $\vp$-direction
circle with momentum
$\nu$   and rotating in a 2-plane with spin $\ss$.
If the  boost energy  is smaller than the rotation one,
i.e. if $ \nu^2 \ll \ss$, then
we are back to the flat-space Regge trajectory \gkp:
\eqn\regg{
\E \approx \sqrt{2 \ss}   + { \nu^2 \ov 2 \sqrt{2 \ss} }
\ . }
A more interesting   case is that of  large $\nu\gg 1$.
Then from \etash\ $\nu \gg 2 \ss$,  and thus \Eshort\
becomes
\eqn\goo{
\E \approx  \nu +\ss +{\ss\ov 2\nu^2} + ...  \ , \ \  \ {\rm i.e. }
\ \ \ \ \ \
E\approx  J   + S   +  {\l   S \ov 2 J^2 } + ... \ . }
Remarkably, this
expression can be directly related to
the leading quantum term  in the
spectrum of strings in \adss  in the frame  boosted
to the speed  of light along the $\vp$-circle of $S^5$ \bmn:
 \eqn\mall{
E= J +
\sum^\infty_{n=-\infty} \sqrt{1+{\l  n^2\ov  J^2 }} \ N_n
\  +\    O({1\ov \sql}) \  . }
Here $J = \sqrt \l \nu $  is the quantum value
of the momentum  and $N_n$  stands for  the oscillator
 occupation number of a particular state
(see  \mets\  for detailed expressions; see also section 4 below).
Let us first recall  that in the flat-space
string theory the state with  a large   angular momentum
on the leading Regge trajectory  can be described
 either  by expanding near
a classical rotating  string configuration or
 by expanding near a point-like string vacuum state
 and building up the angular momentum out of quantum oscillator
 modes (the semiclassical configuration
   with large angular momentum will then
 be represented  by  the corresponding coherent state
 %NN
 $e^{\sqrt S a^\dagger_1} |0>$).
 In the light-cone  frame   where
 $ P^- = - { 1 \ov  \a' p^+} \sum_n  \ |n|  N_n $
 the leading Regge trajectory  $ P^- \sim S$
  will be represented by states
 $ (a^\dagger_1)^S | 0>$   for which  $n=1, \ N_1= S$.

In exactly the same fashion,
the classical energy  \goo\  of rotated  string
in the  frame boosted along the
direction $\vp$ transverse to the rotation plane
is  correctly captured by the quantum
 spectrum \mall\
of string oscillations  in the  $S^5$-boosted frame:
applying \mall\ to the  oscillator state
with  $n=1, \ N_1= S$  ($ 1 \ll S \ll J$)
and expanding in  large $J$
one reproduces \goo.

Note that the  term linear in $S$ in \goo\
comes  from the second term in \ES\ and \Eshort,
i.e. its presence is due to the curvature of \ads.
Indeed, this term is reproduced  by the first
term (``1'') under the square root in \mall\
(related to the  mass term  of the \lc  string
coordinates in the  \pw background \mett)
 whose
origin  can be traced to
the curvature  of the \ads background
\refs{\blau,\bmn}.

This  observation
 supports   the  suggestion  of \gkp\
 that  parts of semiclassical \adss string spectrum
  can be captured by expanding  near  different classical
 string solutions. We see that there is an overlap
 between the leading-order (large $\sql$)
  quantum spectrum   obtained
 by expanding near $S^5$-boosted  point-like string state
 with {\it no}  rotation in \ads and  a
 classical spectrum  obtained by expanding
 near a highly boosted  {\it and} rotating
 string solution.

 %%%%%%%%%%%%%%%%%%%%%%%%%%%%%%%%%%%%%%%%%
 \subsec{\bf Long  string}
%%%%%%%%%%%%%%%%%%%%%%%%%%%%%%%%%%%

 For long strings for which the maximal value
$\r_0$ of $\r(\s)$ is large, i.e.   $\eta \ll 1$
we get from \jjei,\too\  and \ssei
\eqn\kaplo{
\k^2\approx \nu^2+ {1\ov \pi^2}\ln^2 {1\ov \eta}\ ,\ \ \ \ \ \ \ \ \
\oo^2 \approx  \nu^2+ {1\ov \pi^2}\left(1+\eta\right)\ln^2 {1\ov \eta}\
,}
\eqn\sslo{
\ss \approx  {2\oo\ov \eta\ln {1\ov \eta}}  \ . }
We learn that  for long strings
 the  spin is  always large, $\ss \gg 1$,
 irrespective of the value of $\n$.
 Here  there is no simple relation between  $\E$ and $\ss$
 for general $\n$ (cf. \Eshort)   so we need to consider special cases.

 In the case  of small $\n$, i.e.
$\nu \ll  \ln {1\ov \eta}$,  \sslo\  implies that
\eqn\spp{ {1\ov \eta} \approx {\pi \ov 2}\ss\ , }
and  using the relation \ES, we obtain the energy
\eqn\Elongi{
\E \approx  \ss +{1\ov\pi}\ln\ss \ +\   {\pi\nu^2\ov 2\ln\ss } \ ,
\ \  {\rm i.e.} \ \ \ \ \
E \approx   S  +{\sql \ov\pi}\ln {S \ov \sql} \ +\
{\pi J ^2\ov 2 \sqrt{\l} \ln { S \ov \sql}  } \ .
}
We have  kept only the first
$\ln \ss$ -correction and the
leading term in $\nu^2$.
 The first two terms in
\Elongi\ are  the ones obtained in \gkp\
where $\nu$ was equal to zero.\foot{Note
that using  the
relation  \ES\ to find the energy as a function of  spin it is
 sufficient to determine  only the leading
$\ss$-dependence of $\eta$.}

In the opposite case of large $\nu$, i.e.
 $\nu \gg   \ln {1\ov \eta}$ we  find
 from  \sslo,\kaplo\
\eqn\eqeta{
\eta \ln {1\ov \eta} \approx  {2\nu\ov \ss}\ .}
Since $\eta \ll 1$, this  implies that $ \nu \ll  \ss$.
The leading
asymptotics for the solution of the equation  \eqeta\ for  $\eta$ is
\eqn\lead{
 \ln {1\ov \eta} \approx  \ln  {\ss\ov \nu}\ .  }
Using  \ES, we  then find that the  energy
in this case of \  $   \ln {\ss\ov \nu } \ll  \nu \ll  \ss $ is
given by
\eqn\Elongii{
\E \approx  \ss + \nu + {1\ov 2\pi^2\nu}\ln^2  {\ss\ov \nu}
\ , \ \ {\rm i.e.}
\ \ \ \ \
E  \approx  S  +  J + {\l \ov 2\pi^2 J }\ln^2  {S\ov J } \ .
   }
In contrast to the large $\nu$ limit  of the
short string case \goo\ here  the third correction term
is not related to the ``plane-wave'' spectrum \mall.
This is not surprising since there one considered  the limit
of large boost and small string oscillations, while
in the long-string case the spin is always larger than the boost
parameter $\nu$.

Compared  to  the small boost  $\nu$ case
\Elongi\  the energy  $\E$ in  \Elongii\  no longer has a
 characteristic  large-spin
 $\ln \ss$ term found in \gkp.
 In general, $\E(\ss,\n) $ contains \  $k(\n) \ln \ss$ \ term
 with coefficient function  such that
 $ k(\n \ll 1  ) \to  { 1 \ov \pi} $   (cf. \Elongi) and
 $ k(\n\gg 1 ) \to  {\ln \nu \ov \pi^2 \nu} \to 0$
 \ (cf. \Elongii).
 Eq. \Elongii\ predicts the presence of the
 ${\l \ov 2\pi^2 J }\ln^2  {S\ov J } $
 term in the  strong-coupling  asymptotics
 of the anomalous  dimension  of the
 gauge-theory operators   with large  spin and large
  $R$-charge \bmn.
 It would be very  interesting to
 see if the $\ln S$ term
 \gkp\  in the anomalous
 dimensions of the  corresponding
 $N=4$ SYM  operators  with
 large R-charge
 is  suppressed  also at weak 't Hooft coupling.

%%%%%%%%%%%%%%%%%%%%%%%%%%%%%%%%%%%%
\newsec{ 1-loop approximation:  boost  in $S^5$  }
%%%%%%%%%%%%%%%%%%%%%%%%%%%%%%%%%%%

Our aim is to  compute the  leading quantum corrections
to the energy spectrum
by expanding the action
to quadratic order  in fluctuations
near the classical solution.
In flat space,
where the action is gaussian in the conformal gauge,
 this would effectively account for the full string spectrum
of states, irrespective of a
particular  classical solution one starts with.
% (in the sector with fixed
%energy and angular momentum numbers).
In the present non-linear sigma model
 case this will be a good approximation
to the string spectrum  only  up
 to $\aa\ov R^2 $=${1 \ov \sqrt \l}$
corrections.\foot{ As usual, the semiclassical  expansion will
be well-defined if the parameters of the classical solution
($\k,\w,\n, ...$) are
fixed  in the limit ${1 \ov \sqrt \l}\to 0.$}

In this section
we shall  consider  the case of  rotation in $S^5$ only,
i.e. the solution \sll\ with  $\w=0$, \   $\nu=\k$, \ $\r=0$,
reproducing   the  ``plane-wave''  string oscillator
spectrum  of  \bmn\
by an  explicit  version of the  argument suggested in
 \gkp.
We shall use
the \adss string sigma model action in the conformal
gauge.
Given the solution
\eqn\roo{
t= \vp = \n  \tau \ , \ \  \r=0 \ , \ \ \ \ \ \
\beta_{l}=0 \ \  (l=1,2,3) \
,\ \ \ \ \ \ \psi_s =0  \ \  (s=1,2,3,4)\ , }
we  will
expand   the action and the
constraints up to quadratic order
in fluctuations.

Since  the  point $\r=0$ is special
with regard to small fluctuations along $S^3$
directions in \ads
(the corresponding tangent-space metric is degenerate)\foot{This
is the same as starting   with the model
$L = (\del \r)^2  + \r^2 (\del \vp)^2 $ and trying to do
semiclassical expansion near the point $\r=0$.}
it is useful to replace  $(\r,\b_i)$  by
 4 cartesian  coordinates  $\eta_k$
($k=1,2,3,4$)
and then consider their fluctuations.
Equivalently, we
may start with the \ads  metric  \add\ written in the coordinates
\eqn\kop{
ds^2
= - { (1+r^2)^2  \ov (1-r^2 )^2  }  dt^2 +  {4\ov  (1-r^2 )^2 }
 ( dr^2 + r^2 d\Omega_3) =
-  { (1+\fo \eta^2 )^2 \ov (1-\fo \eta^2 )^2  }  dt^2 +
  {  d\eta_k d\eta_k   \ov  (1- \fo  \eta^2   )^2 }  \ , }
 and  then  expand \gss\ near
$\eta_k=0$, i.e.
\eqn\flu{
t= \n  \tau + \kk   \td t  \ , \ \ \
\ \ \ \eta_k=\kk
\td \eta_k  \ ,
\ \ \ \ \ \
\vp = \n  \tau + \kk   \td \vp   \ , \ \ \
 \ \ \
\psi_s = \kk   \td \ps_s  \ . \ }
This leads to  the following bosonic  action
for the quadratic fluctuations:
\eqn\bos{
I^{(2)}_B = - { 1 \ov 4 \pi}  \int d^2 \xi \ \big[
-\del_a \td t \del^a \td t + \del_a \td \vp \del^a \td \vp
+  \n^2 (\td \eta^2_k + \td \psi_s^2)
+ \del_a \td \eta_k  \del^a \td \eta_k  +
 \del_a \td \psi_s \del^a \td \psi_s  \big] \ .
}
This  is the same action that is found  by expanding  the
string  action
in the \pw background of \blau\
\eqn\boso{
I^{(pw)}_B = - { 1  \ov 4 \pi}  \int d^2 \xi \ \big[
\del_a x^+ \del^a x^-
- \fo ( \eta^2_k +  \psi_s^2) \del^a x^+ \del_a x^+
+ \del_a \eta_k  \del^a  \eta_k
+ \del_a \psi_s \del^a  \psi_s  \big] \ ,
}
to quadratic order
near the following  classical solution
\eqn\claas{ x^+ = p^+ \tau \ ,\ \ \ \ \
x^-=0 \ , \ \ \ \ \eta_k=0\ , \ \ \ \  \psi_s=0 \ ,
\ \ \ \ \ \   x^\pm \equiv
\vp \pm t \ , \ \ \    \ p^+ \equiv   2 \n  \ .}
Starting with the
action \boso\ (supplemented with fermions \fer\ to preserve conformal
invariance at the quantum level)
one  may then find the corresponding
spectrum  by
imposing
the quantum light-cone  gauge condition
\eqn\ter{ \td x^+ =0 \  , \ \ \ \ {\rm i.e.} \ \ \ \ \
 \ \ \ \
\td \vp +  \td t =0    \ .  }
The same  is effectively
possible (to the leading order
in $1\ov \sql$ expansion)
 also in the present case,
even though the full  non-linear action \gss\ does not allow
one to
fix the
 \lc gauge in its standard form.
The resulting \lc gauge action is  then simply that of 8 massive
2-d fields as in the case of \boso\ in the \lc gauge
 \refs{\mett}
\eqn\bosy{
I^{(2)}_B = - { 1 \ov 4 \pi}  \int d^2 \xi \ \big[
 \del_a \td \eta_k  \del^a \td \eta_k  +
 \del_a \td \psi_s \del^a \td \psi_s
+ \n^2 (\td \eta^2_k + \td \psi_s^2)
 \big] \ .
}
The  relevant  quadratic  part of the fermionic action \fer\
also  takes a  simple form (the only non-vanishing $\del_a X^M$
factors in \hop\ are  $\del_0  t=\del_0 \vp =\n$)
and becomes  the same as in
\refs{\mett} if we choose the \lc kappa-symmetry gauge
$\Gamma^+ \theta^I=0$ (see  \mets\ and section 5.3   below).
As a result, the quadratic fluctuation
part of the superstring action expanded
near the  solution \roo\  takes indeed
 the same form as the full GS action
in the maximally supersymmetric  \pw background.
Therefore,
 its spectrum  (to the leading order in $1\ov \sql$)
is  the same as found  in   \refs{\mets,\bmn}.

In more detail,  the action \bos\  should be supplemented by
the  conformal gauge constraints \conn,\coonn\ which,
expanded  to quadratic order in the
fluctuations, take the form
$$
 2 \sqf  \n   \del_0 \td x^-
-   \n^2 (\td \eta^2_k + \td \psi_s^2)
+ \del_0 \td \eta_k  \del_0 \td \eta_k  +
\del_1 \td \eta_k  \del_1 \td \eta_k +
 \del_0 \td \psi_s \del_0  \td \psi_s
 +  \del_1 \td \psi_s \del_1 \td \psi_s  $$
\eqn\jjj{ + \  \del_0 \td x^+ \del_0 \td x^-  +
 \del_1 \td x^+  \del_1 \td x^-  + O( \isql) + ...
 =0 \ , }
 \eqn\jjkj{
 2  \sqf  \n  \del_1 \td x^-   +
 \del_0 \td \eta_k  \del_1 \td \eta_k  +
 \del_0 \td \psi_s \del_1  \td \psi_s
  +   \ha  \del_0 \td x^+ \del_1 \td x^-
+ \ha  \del_1 \td x^+ \del_0 \td x^- + O( \isql)  + ... =0 \ ,  }
where dots stand for the   fermionic contributions.
As usual, one may either  impose the  constraints
on  states  ``on average'' or
%SF
impose a proper gauge condition (light-cone
gauge  in the present case),
%SF
directly solve constraints  and quantise the remaining
degrees of freedom. Both approaches lead to the
same expression for the
physical  spectrum,  but the second is more direct
 in the present case.
 Here we should solve constraints  perturbatively
 in  large $\l$, so that we learn that $\del_{a} \td x^-
 \ap 0$;  that means that quadratic
terms
%SF
in $\del \td x^\pm$
%SF
in \jjj,\jjkj\  and  in the action
can be omitted to the leading order. Since $\td x^-$
is ``conjugate''  to $\td x^+$ in the action \bos,
this is essentially equivalent to imposing the quantum \lc gauge
\ter. We thus find the expressions for  $\del_0 \td x^- $
and $\del_1 \td x^- $  in terms of  8 fluctuations
$\td \eta_k, \td \psi_s$.

To  find   the correction to the  space-time energy
conjugate to the  time coordinate $t$  we   note that
to second order in fluctuations the relations  for $P_t$ and
$P_\vp$  are (cf. \eee\ and \kop,\add,\ade)
\eqn\eeee{
 E= P_t=
  \int^{2\pi}_0 {d \s\ov  2 \pi} \ (
  {\sql }
 \n  + \l^{1/4} \del_0 \td t +
     \n  \td \eta_k^2 + ...)
\ , }
  \eqn\jje{
 J= P_\vp=
  \int^{2\pi}_0  {d \s\ov  2 \pi}  \  (
  {\sql } \n  + \l^{1/4} \del_0 \td \vp
      -  \n  \td \psi^2_s + ...)
  \ . }
  %NN
  Dots stand for fermionic and higher order terms.
  Taking their difference  and using the  constraint
  \jjj\  to eliminate the $\del_0 (\td t -\td \vp)$
  in terms of the transverse oscillators
 we find  that  the
extra  terms in  \eeee,\jje\   effectively  change
the sign of the  $\td \eta_k $   and    $\td \psi_s$
 mass terms in the constraint \jjj\
written in the form  $\int d\s \del_0 \td x^-= ...\ $.
%An alternative procedure is to  consider the expectation values
%of \eee,\jje\ in a particular state and to use the
%expectation values of the constraints \jjj,\jjkj\
%to eliminate  $< \int d\s \del_0 \td x^-> $.\foot{Note that just as in
%flat space
% the expectation values of $\del \td x^+ \del \td x^-$ terms
%in \jjj,\jjkj\  do not contribute in this case.}
The end result is  that $E-J$   is  given by  the
expectation  value of the transverse Hamiltonian
(see also Appendix). This is  the same expression
as   found by directly
applying the quantum \lc gauge \ter,
giving  the expectation
value of $-P^-= E-J$
in terms  of the eigen-values  of the
``\lc'' Hamiltonian, i.e.
equivalent (once the fermions are included)
to the  one \mall\  in  \refs{\bmn,\mets}
\eqn\expa{
  E-J =
{ 1 \ov \n } \sum^\infty_{n=-\infty}  \sqrt{ n^2  +  {\n^2 }  }\   N_n \
  + \ O( {1\ov
\sql})   =     \sum^\infty_{n=-\infty}
\sqrt{1 + {\l n^2 \ov J^2} }\  N_n
\ + \ O({1\ov  \sql})  \ . }
Here $N_n$ stands for the  occupation number of the
8 sets of the  bosonic and fermionic oscillators  in
the corresponding oscillator string  state.
Note that this   expression   is
valid only
to  the leading order in  expansion in  $1\ov \sql$;
that     is   why one  is able to
replace  the tree-level  value of the momentum
$\n $ in the first equality in  \expa\
by the exact  expression
$J\ov \sql$.

To clarify the meaning of  \expa\  let us first note that
we are considering a sector of string states  with
given (quantized)  value  of the expectation  value of
the quantum momentum  operator $\hat J$, i.e. for any
state $|\Psi\rangle $ in this sector
\eqn\avv{
  \langle\Psi | \hat J | \Psi\rangle  \equiv  J = J_0 \equiv  \sql \n  \ . }
This should be true, in particular,  for the ground state $|0\rangle$
containing no  transverse oscillations.  An apparent contradiction
with the presence of corrections  in \jje\  is resolved
by noting  that $|0\rangle$ will, in general, be different from the
transverse oscillator vacuum state by terms
 of order $O({1\ov  \sql})$,  i.e.
 $|0\rangle = |0\rangle_0 + {1\ov  \sql} |0\rangle_1 + ... $,
 so that ${}_0\langle 0| \hat J |0\rangle_0 = J_0 + { c_1 \ov \sql }  +
 ....$.
 Then
eq. \expa\  gives the energies of the string
 states with $J= J_0$  which  are  ``close''  to the
  ``ground state''  with $E=J$.  The energy of the ground
 state (which  is a BPS state)
  does not receive a 1-loop correction:
   the vacuum energy contributions
   of the bosonic and fermionic
 oscillators cancel out  because of the effective
  2-d supersymmetry
 \refs{\mett,\bmn,\mets}.

To compute the $O({1\ov  \sql})$
term in \expa\ one is to
go beyond the leading quadratic approximation. This  gives
 a non-linear action containing
quartic  and higher terms in
 $\td t,\td \vp, \td \psi_s, \td \eta_k$
 (its explicit form follows from the
 metric of $S^5$ \add\  and the metric of \ads  \kop).
 To quartic order in the fields
$$
I^{(4)}_B = - { 1 \ov 4 \pi}  \int d^2 \xi \ \bigg[
- (1 + \isql \td \eta^2_k )  \del_a \td t \del^a \td t
+  \n^2 [ \td \eta^2_k    + { \hal }\isql (\td \eta^2_k)^2  ]
$$  $$
+\  (1 + \isql \td \psi^2_s )
 \del_a \td \vp  \del^a \td \vp
+  \n^2 [\td \psi_s^2  +  \isql (
\tri   \td \psi_s^4 +
\sum^4_{1=s< s'}\td \psi_s^2 \td \psi_{s'}^2 ) ]
$$
\eqn\gyg{
+\   (1 + \hal \isql \td \eta^2_k)
\del_a \td \eta_n \del^a \td \eta_n
+ \sum^5_{i=1} ( 1 - \isql \sum^{i-1}_{k=1}\td \psi_k^2 )
 \del_a \td \psi_i \del^a \td \psi_i \bigg]
\ . }
Here  we  may again    proceed with solving the  corresponding
conformal-gauge   constraints perturbatively
in $\isql$. This  will be
effectively  equivalent  to imposing
the \lc gauge  at lower order in $\l$, i.e.
to ``deforming''  the gauge condition $x^+ =p^+ \ta + O(\isql)$.
One would also need to take
into account  the quartic fermionic terms  in  \fer.

%N
It is possible  to  determine the general structure of
higher-order string sigma model  loop ($\a' \sim {1\ov \sqrt \l}$)
corrections to the energy relation  \expa.\foot{We are grateful to D. Berenstein
and I. Klebanov  for important
 discussions of this issue.}
Using that the sigma-model action
\gss\ expanded near the classical
solution
\roo\  should  contain
 only two derivatives or two powers of $\nu$
in all  interaction terms (cf. \gyg) and  should lead to
UV finite  quantum 2-d effective action one concludes
that $l$-loop correction to the  space-time
 energy  of a particular
oscillator string  state  with quantum  numbers
$n_i$ should have the  structure
$(\Delta E)_l=
{1 \ov \nu} (\Delta E_{\rm 2-d})_l =
({1 \ov \sqrt \l\ \nu })^{l-1} F_l ( {1\ov \nu^2}; n_i) $.
The function $F_l$ should vanish  for $n_i=0$
(the ground state energy should not receive corrections on
 the basis of supersymmetry) and should  have a
 regular expansion in positive powers of $1\ov \nu^2$,
 approaching  a constant  for $\nu \to \infty$.
Expressed in terms of
$J= \sqrt \l\ \nu$  this gives:
$(\Delta E)_l=
{1 \ov J^{l-1}} F_l ( {\l \ov J^2}; n_i) $,
with $F_l$  having power series expansion in ${\l \ov J^2}$
(cf. \expa\ for $l=1$). Thus, in the large $J$  sector
one is able to re-interpret the string $\a'\sim {1\ov \sqrt \l}$
expansion  as expansion  in positive  powers
of ${\l \ov J^2}$,  opening an interesting
 possibility of  direct
comparison  with perturbative  computations  of anomalous
dimensions  of corresponding operators  with
large R-change on the SYM  side.

%%%%%%%%%%%%%%%%%%%%%%%%%%%%%%%%%%%%
\newsec{ 1-loop approximation: rotation in \ads  and boost  in $S^5$  }
%%%%%%%%%%%%%%%%%%%%%%%%%%%%%%%%%%%

We  would like  now to repeat the   analysis of the previous
section  in the case
of a more general solution \sll, including rotation in \ads.
The  goal is to determine  the  quantum
 string spectrum
at the leading order  in  large $\sqrt \lambda$.
We shall   expand the string action
 near the background \sll\  as in    \flu:
$$
t= \om  \tau + \kk   \td t  \ , \ \ \
\ \ \ \ \ \ \r= \r(\s) + \kk   \td \r \ ,
\ \ \ \ \
\
\p = \w  \tau + \kk   \td \p   \ , \ \ \ \ \
\vp = \n \tau +   \kk   \td \vp   \ ,   $$\eqn\exx{
\beta_i  = \kk   \td
\beta_i  \ \ \ \  (i=1,2) \ , \ \ \ \ \ \ \ \
\psi_s = \kk   \td \ps_s  \  \ \ \ \ (s=1,2,3,4)\ .  }
As in the previous section,
one possible strategy  is to use
the conformal gauge  and  expand the action and  the
constraints to the second order in the fluctuations,
solve the  constraints  perturbatively  in $1 \ov \sql$,
express the energy $E$ in terms of the  quantum fields
and then  compute the  leading-order
quantum  term in the spectrum.

As we shall see below,  this
 conformal gauge  approach   turns out to be more
 complicated than in the pure  $S^5$ boost  case:
 in contrast to the  discussion
 in the previous  section  where the two  fluctuation fields
 $(\td t, \td \vp)$ effectively decoupled (cf. \bos),
  here $\td t$ will mix   non-trivially
   with $\td \p$ and $\td \r$. Also, the masses of
  the fields
  will depend on $\s$, making it hard to find the spectrum
  exactly. A study of the
  resulting  1-loop correction to the energy spectrum
   will be the subject
 of  section 6.

 An a priori  possible alternative to the conformal gauge
 approach is to start
 with the Nambu action  in  the  static gauge, i.e.
 to consider only fluctuations  normal to the embedded 2-d
 surface.
 However, this approach  turns out to be
  problematic  in the present case where the induced metric \coop\
 has singularities.
   While the resulting action
 for small fluctuations is  somewhat  simpler than the
  conformal gauge action,  one of the mass terms
contains a 2-d curvature part  and thus is singular  at the
turning points.
In addition,
the resulting  bosonic+fermionic  action   does not appear
to be
finite:  like in  the case of the 1-loop correction to the
Wilson loop  \refs{\theis,\dgt}
the 2-d logarithmic divergences do not
manifestly cancel out, implying a subtlety \dgt\ in the
 definition of the quantum measure in the
 approach based  directly  on   the  Nambu action.
 As we shall see,
 problem  can be  avoided \dgt\ in the Polyakov approach  to GS string
 in the conformal gauge:  since the conformal anomaly
cancels out in the flat 10-d space,
 one  may  trade the conformally-flat  induced
 metric for  a flat background  metric on the
  cylinder ($\tau,\s$). Then
    the  2-d ghosts decouple, all masses are regular
    for all values of $\s$
  and the finiteness of the 2-d quantum  theory defined by the
   quadratic
  fluctuation action  becomes  manifest.\foot{This is  a
  consequence   of a
  mass-squared sum rule  which itself  is
  a consequence
  of the effective 2-d supersymmetry (spontaneously broken
  by the classical background).}

 Below we shall start with  computing the
 bosonic part of the
 action,  both    in the static  gauge  and
  in the conformal gauge,
 and then  determine  the  corresponding
 quadratic fermionic term.

%%%%%%%%%%%%%%%%%%%%%%%%%%%%%%%%%%%%%%%%%
\subsec{\bf  Bosons:  static gauge  }
%%%%%%%%%%%%%%%%%%%%%%%%%%%%%%%%%%%%%

Let us start with the Nambu action  and impose
 the static gauge on $t,\r$, i.e. set their fluctuations to zero:
\eqn\sta { \td t =0 \ , \ \ \ \ \ \ \
\td \r =0 \ . }
The  form of the resulting action for small fluctuations
is then a special case  of the general expression
for a semiclassical expansion  of the  string action in \adss
(see \refs{\fro,\theis,\dgt}).
 For  $\nu=0$
one finds  one ``transverse''
scalar field  fluctuation $\bar \p$  with a non-trivial 2-d curvature
dependent
mass term,
  2 massive fluctuations in the two other
  directions
of $S^3$ and 5 massless $S^5$ fluctuations
%For generic value of $\nu$ we get
 \foot{The same  expression
for the quadratic-fluctuation action is found
by expanding not in  $\td \p$,  but in  the direction
$ \zeta = \r'^{-1} ( \w \tanh \r \ \td \t + \k \coth \td \p ) $
normal to the embedded 2-d surface.}
 $$
I^{(2)}_B
=  - { 1 \ov 4 \pi}  \int d^2 \xi \sqrt{-g}  \big[
g^{ab}   \del_a \bar  \p  \del_b \bar \p  +
  (4 +   R^{(2)})
  % + \n^2 g^{00}
      \bar \p^2
  +
g^{ab}  \del_a \bar \b_i  \del_b  \bar \b_i   +
 2
    \bar\b_i^2  $$
 \eqn\stat{+  \
 g^{ab}  \del_a \td \vp  \del_b  \td \vp
   + \
  g^{ab}  \del_a \td \psi_s  \del_b  \td \psi_s
   \big] \ ,
}
where $g_{ab}$ is the induced metric  \coop\
and  $R^{(2)}$ is its curvature
\cuu.
The fields $\bar \p$ and  $ \bar \b_i$
are related to $\td \p$ and  $ \td  \b_i$ in \exx\
by a $\r$-dependent rescaling  needed  to put the
kinetic terms  in the canonical form.
In more detail,
expanding the square root of the Nambu action we get
the following term  depending  on $\td \p$:
\eqn\jui{
L (\td \p)  = -\r'^2  + \ha \sinh^2 \r\ (1   +
{ \w^2 \sinh^2\r \ov \r'^2})
[  (\del_0 \td \p)^2  - (\del_1 \td \p)^2 ] + O(\td \p^4)
\ ,  }
where $\r'$ is given by \rgo.
After the  field redefinition that  puts the kinetic term
in the canonical form
\eqn\hoh{
\bar \p  \equiv  { \sinh\r \ov
\sqrt{ 1 - { \w^2 \ov \k^2 } \tanh^2 \r} } \td \p
 \ , }
we get the $\bar\p$ part of \stat.
Explicitly, using that the induced metric is
conformally flat \stat\ can be written as
\eqn\stit{
I^{(2)}_B
=  - { 1 \ov 4 \pi}  \int d^2 \xi   \big[
  \del_a \bar  \p  \del^a  \bar \p  +
  m^2_\p  \bar \p^2
  + \del_a \bar \b_i  \del^a  \bar \b_i   +
 m^2_{ _\b }  \bar\b_i^2   +  \del_a \td \vp  \del^a   \td \vp +
   \del_a \td \psi_s  \del^a   \td \psi_s \big] \ ,
}
where (using \cuu)
\eqn\jou{
m^2_\p =
\sqrt{-g}  (4+  R^{(2)})   =
2 \r'^2  +    { 2 \k^2 \w^2 \ov  \r'^2} \  ,  }
\eqn\mos{  m^2_{_\b} = 2 \r'^2 =
2 [ ( \k^2-\w^2)  \cosh^2\r   + \w^2  ]  \ . }
As we shall see below in section 5.3,
for $\n=0$ the contribution of the GS fermions
can be effectively
represented by  8 copies of 2-d Majorana
fermions  with masses  $\pm 1 $
defined on  the 2-d surface with induced metric,
i.e.  with the same
 squared Dirac operator $- \hat \nabla^2 + \four
R^{(2)} + 1$ as in \refs{\dgt,\theis}
($\hat \nabla$ is the covariant derivative
in the induced metric).
 As in the Wilson loop case considered in   \theis, the
  logarithmic divergences  coming from
  the fermionic part of the action
  will not then   cancel against   the
  divergences  corresponding to the  2-d  bosonic
  theory \stit\
   \eqn\hyh{ b_2=
   \sum_k \sqrt{-g}( { R^{(2)} \ov 6 } d_k   - m^2_k
   )  =
 \sqrt{-g} \big[{ R^{(2)} \ov 6 } ( 1 + 2 + 5)   - ( 4 +
 R^{(2)} + 2 \times 2) \big] \ .   }
 As discussed in \dgt, the remaining
 divergence $\sim R^{(2)}$  should be  cancelled
 out  by a measure contribution  needed  to make the  Nambu
 string  partition function
 equivalent  to the partition function in
  the Polyakov approach (with the induced metric used
  as the  background one).

 There is also  another apparent
 problem related with the presence of that
 $R^{(2)}$ term in the mass  \jou: this term  blows up  at the
 turning points, while all other masses  go there to zero.\foot{
 The vanishing of masses at the turning points is  an expected
 behaviour as  the fold points  should
 move with velocity of light.}
 Both  problems can be avoided  by  starting with
  the Polyakov  action  in the  conformal  gauge
  where (due to cancellation of conformal anomaly)
  one is able to choose  the background metric
  on the cylinder  to be flat, making
  all masses regular  at the folding points.

Let us mention that  for non-zero $\n$
the action \stat\ becomes more complicated:
the  fluctuations  $\bar \p$ and $\td \vp$ mix in  a non-trivial
way. This mixing will be absent in the  conformal gauge
where the \ads and $S^5$ parts of the  bosonic action
are decoupled (but it will reappear if one would  try to solve
the constraints  and use their solution  in the action).

%%%%%%%%%%%%%%%%%%%%%%%%%%%%%%%%%%%%%%%%%
\subsec{\bf Bosons: conformal gauge  }
%%%%%%%%%%%%%%%%%%%%%%%%%%%%%%%%%%%%%

Let us  now  present  the expression for the
quadratic fluctuation action
in the conformal gauge.
Expanding the sigma model as in
\flu,\exx\ gives
  the following bosonic  action
for the quadratic fluctuations (cf. \bos)
  $$
I^{(2)}_B = - { 1 \ov 4 \pi}  \int d^2 \xi \ \big[
-\cosh^2 \r \ \del_a \td t \del^a \td t
+ \sinh^2 \r \   \del_a \td \p
\del^a \td \p    + \ 2 \sinh 2\r \ \td \r ( \om \del_0 \td t
-  \w \del_0 \td \p ) $$  $$
+  \ \del_a \td \r \del^a \td \r
 +  ( \k^2-\w^2)  \cosh 2 \r \ \td \r^2
 + \ \sinh^2 \r\  (
\del_a \td \b_i  \del^a \td \b_i   +  \w^2\td  \b_i^2 )
 $$  \eqn\dobb{
 + \ \del_a \td \vp \del^a \td \vp
+  \del_a \td \psi_s \del^a \td \psi_s   +   \nu^2
 \td \psi_s^2 \big] \ .
}
The linearised  (leading-order in $1\ov \sqrt{\l}$)
part of the  conformal-gauge constraints  \conn,\coonn\
corresponding to the action \dobb\   has the form
(cf. \jjj,\jjkj)
 \eqn\constri{
 -\k \cosh^2\r\, \pa_0\tilde{t} + \oo \sinh^2\r\,
 \pa_0\tilde{\phi}
+ \nu \pa_0\tilde{\varphi} + \r' \pa_1\tilde{\r}+ {1\ov 2}(\oo^2 -
\k^2)\sinh 2\r \, \tilde{\r} \ap  0
\ ,}
\eqn\constrii{-\k \cosh^2\r\, \pa_1\tilde{t} + \oo \sinh^2\r\,
\pa_1\tilde{\phi}
+ \nu \pa_1\tilde{\varphi} + \r' \pa_0\tilde{\r} \ap  0\ .}
Using these constraints to  eliminate one
of the  fluctuation fields,
e.g.,  $\td t$  (cf. $\td x^-$ in  \jjj,\jjkj)
 from the action \dobb,   one should discover
 that  some linear   combination
  of the remaining  fields (cf. $\td x^+$ in  \bos)
then  decouples from the action (it should then  be formally
 gauge-fixed to zero  to avoid degeneracy of the path integral).
 The resulting action  for  the 8 ``transverse''
   bosonic
 fluctuation fields  should be (classically)
 equivalent to the  static-gauge  action  \stit.

Redefining  the fields to put the  kinetic terms in the
standard form
\eqn\rede{
\bar t = \cosh  \r \  \td t \ , \ \ \
\bar \p  = \sinh \r \  \td \p \ , \ \ \
\bar \b_i   = \sinh \r \  \td \b_i  \ , \ \ \
\ \bar \r \equiv \td \r\  ,
}
transforms   \dobb\ into
$$
I^{(2)}_B \equiv I_1( \bar t, \bar \p, \bar \r) + I_2
( \bar \b_i ,\td \vp, \td \psi_s)   \ ,
$$
$$ I_1
=  - { 1 \ov 4 \pi}  \int d^2 \xi \ \big[
-  \del_a \bar t \del^a \bar  t   - \m^2_t \bar t^2
+   \del_a \bar  \p  \del^a \bar \p  + \m^2_\p \bar \p^2
 $$  \eqn\onn{
+ \ 4  \rr ( \om \sinh \r \ \del_0 \bar t
-  \w \cosh \r \ \del_0  \bar \p )
+   \del_a \rr \del^a \rr   + \m^2_\r  \rr ^2 \big]
\ , }
   \eqn\obb{ I_2 =
   - { 1 \ov 4 \pi}  \int d^2 \xi \  \big[
 \del_a \bar \b_i  \del^a \bar \b_i   + m^2_{_\b} \bar\b_i^2
+    \del_a \td \vp \del^a \td \vp
+  \del_a \td \psi_s \del^a \td \psi_s   +   \nu^2
 \td \psi_s^2  \big] \ ,
}
where (using \rgo\  and \secc)
\eqn\maas{
\m^2_t =  \r'^2  + \tanh \r \ \r''  = 2 \r'^2  - \k^2 + \n^2
\  , \ \ \ \ \
\m^2_\p  =  \r'^2  + \coth \r \ \r'' = 2 \r'^2   -    \w^2  + \n^2  \ , }
\eqn\joy{
  \m^2_\r =  (\k^2 - \w^2)  \cosh 2 \r
  =  2 \r'^2     - \k^2 - \w^2  + 2\n^2   \
\ , \ \ \ \ \
    m^2_{_\b} = 2 \r'^2  + \n^2
       \ .}
The non-constant part of all masses is thus the same
and  is equal to  $ 2\r'^2$  (cf. \stit).

While in the case of the pure boost   in $S^5$
the small fluctuation operators  had  constant coefficients
(cf. \bos), here we have a  non-trivial dependence on $\s$ through
the classical solution $\r(\s)$. This is very similar
to the case of the  1-loop expansion \refs{\kal,\theis,\dgt} near
the classical  string
 profile appearing in the  Wilson loop \marey\ calculations.
The system splits into  three non-trivially  coupled  fluctuations
    of the effective $AdS_3$ sector,  two  massive  fluctuations in
$S^3$   and  one massless and 4 massive  fluctuations  in $S^5$.

The   action \onn\ can be obtained  in a more geometrical way
as a special case of
 the general semiclassical expansion
of the \adss  sigma model  in terms of tangent-space
fluctuations  \dgt.
Expanding  the   string
sigma model  action \gss\  about a classical solution
$
x^M \to  x^M_0  + z^M ,$
and  introducing  the tangent-space components of the fluctuation
fields
$
z^A = E^A_M(x_0)  z^M ,$  we
get the following action for the quadratic fluctuations
\eqn\ctoq{
I^{(2)}_{ B} = -{ 1 \ov 4\pi } \int d^2 \xi \sqrt {-g}
\left( \eta_{AB} g^{ab} \D_a z^A \D_b z^B
+ X_{AB} z^A z^B \right),
}
\eqn\yyutrq{
X_{AB} = - g^{ab}  e_a^{\,C} e_b^{\,D} R_{ABCD}\,,
\ \ \ \ \
e_a^{\,A} \equiv \del_a  x^M_0  E^a_M ( x_0) \ .
}
 $\D_a$ is the covariant derivative containing
the projection of the target space spin connection,
\eqn\strcovd{
\D_a z^A = \del_a z^A + w^{AB }_a z^B \,,\ \ \ \
\qquad
w^{AB}_a= \del_a  x^M_0 \omega^{AB}_M \,
,}
where $\omega_M^{AB}$ is the spin connection of
the target space \adss. The $S^5$ part  of \ctoq\
is the same as in \obb,  while for
the \ads part  with   the curvature
$ R_{ACBD }=-\eta_{AB }\eta_{CD}+\eta_{AD}\eta_{CB}$
we get
\eqn\rewqq{
X^{AB} = \eta_{CD} g^{ab} e_a^{\,C} e_b^{\,D} \eta^{AB}
- g^{ab } e_a^{\,A} e_b ^{\,B } \ . }
Since  the induced metric  $g_{ab}$
is a conformally flat \coop\  it
can be replaced by $\eta_{ab}$.
Explicitly, in the global  coordinates  in \add\ we find
by expanding near the solution \sll:
$z^A= ( \bar t, \bar \r, \bar \p, \td \b_i) \ , \ \ \
E^0_0= \cosh \r\ , \ \ E^1_1 = 1 \ , \
E^2_2= \sinh  \r\ ,   \ \ E^i_i = \sinh  \r\ ,  $
\eqn\bbb{
w^{01}_0= \k \sinh \r \ , \ \ \ \
w^{21}_0= \w \cosh \r \ , \ \ \ \ w^{AB}_1= 0 \ , }
$$ e^0_0= \k \cosh \r \ , \ \ \ \ e^2_0 = \w \sinh \r \ ,
 \ \  \    e^1_1= \r' \ ,\ \ \ \
 \eta_{CD}\eta^{ab} e^C_a e^D_b = 2 \r'^2  + \n^2 \ .  $$
 Note that the 2-d  $SO(1,2)$  ``gauge potential''
  $w^{AB}_a$  depends only on $\r(\s)$
  and  has a non-vanishing field  strength.
 The non-trivial ``$AdS_3$'' part of
 the quadratic action  \onn\ is then
$$
I_1  = -  { 1 \ov 4 \pi}  \int d^2 \xi \ \big[
(\del _0  \bar t     + \k \sinh \r\ \rr )^2
- (\del_1 \bar t)^2
- \  (\del _0  \bar \p      +  \w \cosh \r\ \rr )^2
 +   (\del_1 \bar \p)^2   +   $$ $$
-\   (\del _0  \rr   + \k \sinh \r \ \bar t   -
 \w \cosh \r\ \bar
\p  )^2   +  (\r'^2 + \n^2)    \bar \r^2
+   (\del_1 \rr)^2  $$ \eqn\hipo{
+\ ( 2 \r'^2 + \n^2)  ( - \bar t^2 +  \bar \p^2)
 +  ( \k  \cosh  \r\ \bar t -  \w \sinh \r\ \bar \p)^2
\big] \ . }
This is easily seen to be equivalent  to \onn.
Note also that in the case of  no rotation in \ads, i.e.
$\r=0, \ \w=0, \ \k=\n$  the mass of $\bar t$  vanishes
while the masses of $\bar \p$ and $\bar \r$ become equal to $\n$,
in agreement with \bos\ ($\td \eta_k$ in \bos \
corresponds to $\bar \p,\bar \r, \td \beta_i$  here).

%%%%%%%%%%%%%%%%%%%%%%%%%%%%%%%%%%%%%%
In general, the  action in conformal gauge
 should be supplemented by the 2-d ghost action
\dgt\ ($\a=0,1$):
\eqn\gho{
I_{gh} = -  { 1 \ov 4 \pi}  \int d^2\xi \ \sqrt{-g} \
\ (   \nabla^a \zeta^\a \nabla_a \zeta_\a  - \ha  R^{(2)}
\zeta^\a \zeta_\a ) \ , }
where $\nabla_a \zeta^\a = \del_a \zeta^\a + \omega^{\a\b}_a
\zeta_\b$  is the 2-d covariant derivative with respect to
the  background  2-d metric. In the case when
this metric  is identified with the
induced  (conformally flat) metric \coop\ we have:
$g_{ab} = \ee^\a_a \ee^\b_b \eta_{\a\b} $, \
 $\ee^\a_a = \r' \d^\a_a, \  \zeta^\a = \ee^\a_a \zeta^q$
and $  \omega^{01}_0= { \r''\ov \r'} \ , \
\omega^{01}_1=0$.
Since  the total conformal anomaly  must cancel  after the fermions are
included (see discussion in \dgt),   we may  choose
 the background  metric  on the  2-cylinder  (which is used
to define the  norms and operators in the conformal gauge)
 to be
not the induced  but  simply  the  flat  metric.
This is quite natural, given the singularity of the induced metric.
Then  the ghost contribution  becomes
 trivial and can be ignored,
and we are left  with a system
of fields on a flat cylinder  with non-constant
(but everywhere regular) masses.

The resulting  contribution of the fields in
\hipo\ to the  2-d logarithmic divergences
is then proportional to the trace of their mass matrix
(cf. \hyh)
$$ \r'^2 + \n^2
 +   2 \times (2 \r'^2 + \n^2)  - \k^2 \cosh^2 \r +  \w^2 \sinh^2 \r=
   2  (2 \r'^2 + \n^2)  \ . $$
   The total  contribution of the  bosonic fluctuations
   $\bar t, \bar \p, \bar \r, \td \b_i, \vp, \psi_k$
 \eqn\hyhh{ b_2= - \eta^{AB} X_{AB}  =
   -  2 (2 \r'^2 + \n^2)   -   2 \times (2 \r'^2 + \n^2)    - 4 \n^2
 = -  8 ( \r'^2 + \n^2)  \ }
  will be exactly  cancelled  by the contribution of
  the fermions,  checking  the  conformal invariance
  of the theory.
    \foot{For $g_{ab}$ chosen  to be the induced metric,
  the  contributions of the two  ghosts to the partition function
  should
effectively cancel the contributions
of the two ``longitudinal''  fluctuations
(combinations of $\bar t$ and $\bar \r$)
leading to the same final result as in the static gauge.
In particular, the extra ghost  contribution
to the  mass$^2$ term in the
divergences $ \Delta b_2 =\sqrt { -g} (
-  2\times \ha   R^{({2}) }  ) $
reproduces the $R^{(2)}$ term in \hyh\ (see \dgt).
Since the ghost contribution   can be  ignored in the case
of the flat choice of the fiducial metric, this suggests
that the correct prescription for quantisation
in the static gauge should
 be to omit the
$R^{(2)}$  term  in the mass of $\bar \p$ in \jou\
and define the norms of the fields using  flat metric on the
cylinder.
}

%%%%%%%%%%%%%%%%%%%%%%%%%%%%%%%%%%%%%%%%%%%%%%%%%%%%

Let us comment on   some   special cases.
When  $\n=0$
 and $\w \gg \k$    the range of variation of $\r$ is
small  and we should  be
  back to the flat space  case,  i.e.
the fluctuations should diagonalize and become massless.
For  $ \r\to 0,\ \w \gg \k$
\onn\ can be written as
 \eqn\onnp{ I_1
\approx   - { 1 \ov 4 \pi}  \int d^2 \xi \ \big(
-  \del_a \bar t \del^a \bar  t
+  \del_a\bar  \p  \del^a \bar \p  - \w^2  \bar \p^2
-   4 \w \rr   \ \del_0  \bar \p
+   \del_a \rr \del^a \rr   - \w^2  \rr ^2 \big)
\ .  }
  This  should be same as  the action  obtained by
starting with the metric $ ds^2 = - dt^2 + d\r^2 + \r^2 d\p^2$
and repeating the expansion  near $ t=\k \tau, \
\p= \w \tau, \ \r'^2 = \k^2 - \w^2 \r^2$ solution.
Indeed, after the  redefinition\foot{This  effectively
amounts to  going back to cartesian coordinates
$x_1 = \r \cos \p , \ x_2 = \r \sin \p$
and then expanding  near classical  background.}
$
\eta_1 = \cos \w \tau \ \rr - \r \sin \w\tau \ \td  \p
=    \cos \w \tau \ \rr -  \sin \w\tau \ \bar  \p
  , $  $
\eta_2 = \sin \w \tau \ \rr + \r \cos \w\tau \ \td  \p
=  \sin \w \tau \ \rr +  \cos \w\tau \ \bar  \p ,  $
we get  the massless Lagrangian  $\sim
 (\del \eta_i)^2 $.

Next,  let us  check  that the  expansion  near the
point-like solution  with  $\r'=0$, $\r=\r_0$,
$\k=\w=\n$  (particle located at radius $\r_0$
and rotating in both  $S^5$ and  $S^3$) is indeed
 equivalent to the
expansion  near the solution with $\k=\n$, $\w=0$,
$\r =0$. Here   \onn\  becomes
\eqn\onnr{
I_1( \bar t, \bar \p, \bar \r)
=  - { 1 \ov 4 \pi}  \int d^2 \xi \ \big[
-  \del_a \bar t \del^a \bar  t
+   \del_a \bar  \p  \del^a \bar \p
-   4 \k  \del_0 \rr (  \sinh \r \  \bar t
-   \cosh \r \   \bar \p )
+   \del_a \rr \del^a \rr  \big]
\ . }
 Applying first the $SO(1,1)$ rotation
 $ \hat \p =  \cosh \r \  \bar \p -  \sinh \r \  \bar t , \
 \   \hat t  =  \cosh \r \  \bar t  -  \sinh \r \  \bar \p ,
$  and then the above  $SO(2)$ rotation
in the $\eta_i=(\rr,\hat \p) $ plane
we end up with the Lagrangian
$- \del_a \hat t \del^a \hat   t +
 \del_a \eta_i   \del^a \eta_i + \n^2 \eta_i^2   $, which
 is in agreement  with \bosy.

%%%%%%%%%%%%%%%%%%%%%%%%%%%%%%%%%%%%%%%%%%%
\subsec{\bf Fermions}
%%%%%%%%%%%%%%%%%%%%%%%%%%%%%%

The  derivation of the quadratic fermionic action from \fer\
is similar to the one in  \refs{\dgt}.
The  2-d projections  of $\G$-matrices that enter  the
fermionic action  are (the indices $A=0,1,2,9$  are  used
to label the
$t,\r,\p,\vp$ directions in the
tangent space):\foot{Considering the $(t,\r,\p,\vp)$ subspace
we may   introduce the  two
tangent  $t^M_\a =
 \ee^a_\a \del_a  x^M_0$
  and
 2  normal  vectors $n^M_u$  to the
 embedded world sheet,
 $G_{MN} t^M_\a t^N_\b = \eta_{\a\b},
 \
 G_{MN} t^M_\a n^N_u = 0, \
 G_{MN} n^M_u n^N_v = \d_{uv}$
  ($M=0,1,2,9$ counts the
 directions
 $t,\r,\p,\vp$). Then we find that
 $
 t^M_{\hat 0} = \r'^{-1} ( \k,0,\w,\n ) ,
 \ \
 t^M_{\hat 1}   =  ( 0,1,0,0)  , \  $
 and $n^M_2 = \g^{-1}   ( \w \tanh \r,0,\k\coth\r, 0 )
  , \ \
% n^M_2 = \r'^{-1} ( a,0,b, c)\ , \ \
 n^M_9 = \r'^{-1} (\g^{-1}  \n \k ,0, \g^{-1}  \n \w  ,
 \g ) , \  \g^2 \equiv \r'^2 + \n^2 \ .  $ }
$
\vr_\a = \ee^a_\a  E^A_M (x_0) \del_a x^M_0 \G_A
= t^A_\a \G_A $, \ $ \ee^a_\a  = \r'^{-1} \delta^{a}_\a,  $
or explicitly ($\a=\hat 0, \hat 1$)
 \eqn\hra{
\vr_{\hat 0} = \r'^{-1}  (  \k \cosh \r\ \G_0 +  \w \sinh\r\ \G_2
+  \n  \G_9) \ , \ \
\ \ \  \vr_{\hat 1}= \G_1 \ , \ \ \ \ \  \vr_{(\a} \vr_{\b)} =
\eta_{\a\b} \ .   }
Let us first consider the  more transparent
case of $\n=0$.
As in  \dgt\
one can make a local $SO(1,9)$ rotation
which transforms the set of $\s$-dependent
10-d Dirac matrices into  10 constant
Dirac matrices,
$
\vr_\a (\s) = t^A_\a \G_A = S(\s) \G_\a S\inv (\s)  ,
\ \
 \vr_u (\s) = n^A_u \Gamma_{A}
 = S(\s) \G_u S\inv (\s) \ .
$
One
is then able to write the  quadratic part of the GS action \fer\
as an  action for a set of 2-d Dirac
fermions coupled to curved induced 2-d metric.
Explicitly, in the present case
 we observe that
(see \rgo)
\eqn\soo{
\vr_{\hat 0 } = \cosh \a\ \G_0 + \sinh\a\ \G_2 \ ,  }
\eqn\poo{
\cosh \a\ = {\k \cosh \r
 \ov \sqrt { \k^2 \cosh^2\r - \w^2 \sinh^2 \r}}  \ , \ \ \ \ \
 { d\a \ov d \s} =  { \k \w\ov \r'} =
  \sqrt{ \w^2 \cosh^2 \a - \k^2 \sinh^2\a  }  \ ,
 }
so that  the required transformation
$S$ is a boost in the $(t,\p)$  plane:
\eqn\boo{  S= \exp (\ha \a  \G_0 \G_2 ) \ .  }
After the field redefinition
$
\t^I \to \Psi^I \equiv S\inv \theta^I,$
we then find   that \fer,\form\ becomes
\eqn\frmk{
L_F = i\big( \sqrt{-g} g^{ab}\d^{IJ} - \ep^{ab}
s^{IJ}\big)
 \big( \bar \Psi^I \tau_a {\hat  \nabla }_b \Psi^J
 - \ha i \ep^{JK} \bar\Psi^I \tau_a \G_*\tau_b  \Psi^K\big)
\,,}
where $\tau_a$ play the role of
the curved space 2-d Dirac matrices
\eqn\ttrr{
\tau_a = e_a^{\a} \G_{\a} = \r' \d_a^{A}\G_A   \,,
\qquad
\tau_0 = S\inv \vr_0 S =\r'  \G_0\,,
\qquad
\tau_1 = S\inv \vr_1 S = \r'  \G_1\,,
}
and $\hat \nabla_a$ is the 2-d curved space spinor covariant
derivative for the  induced metric $g_{ab} = \r'^2 \eta_{ab}$,
\eqn\gty{ \
\hat \nabla_a =\del_a + \four \omega^{\a\b }_a \Gamma_{\a \b}\,,
\qquad
\hat \nabla_0 = \del_0  + \ha  { \r''\ov \r'} \G_{01} \ ,
\qquad
\hat \nabla_1 = \del_1\, .
}
In more detail,  the  result of  applying the rotation \boo\
 to the
covariant  derivative $\D_a \equiv \del_a + \four \omega^{AB}_a
\G_{AB}$  in \fer, i.e.
\eqn\yuy{\D_0 = \del_0  - \ha \G_1 ( \k \sinh \r\
 \G_{0}  + \w \cosh\r\  \G_{2} )
\ ,  \ \ \ \ \ \ \
\D_1 = \del_1 \ , }
is
$\td \D_a = S^{-1} \D_a S  = \hat \nabla_a + B_a \ , $
\ $B_a \equiv    { \k\w \ov 2\r'}   \G_{2} \  ( \G_1, -\G_0) $,
but since $ \eta^{\a\b} \G_\a B_{\b} =0, \
 \epsilon^{\a\b} \G_\a B_{\b} =0$,
 the extra connection term cancels out in the
 rotated fermionic
  action \fer\  and  we are left with the same result
  \frmk\ as in \dgt.
 One can then  fix the kappa-symmetry gauge
  by imposing, e.g.,  $\Psi^1=\Psi^2$,  thus ending up with
  a similar  action as in \dgt\
  \eqn\fuu{
L_F = 2i \sqrt{-g}
 \left( \bar \Psi \tau^a {\hat \nabla}_a
\Psi   +  i \bar \Psi  M  \Psi \right)
\,,}
\eqn\you{ M  \equiv {\textstyle {\ep^{ab} \ov 2\sqrt{-g}}} \tau_a  \G_* \tau_b
 = \tau_3 \G_* =   i  \G_{234}\ , \ \ \ \ \ \
 \tau_3 \equiv    {\textstyle{\ep^{ab} \ov 2\sqrt{-g}}} \tau_a \tau_b =
\G_0 \G_1\,,
\qquad   M^2 =1 \ .
}
Choosing a  representation for $\G_A$ such that
$\G_{0,1}$ are essentially  2-d Dirac matrices times
 a unit $8\times 8$
matrix and $\tau_*$ is diagonal, we end up with 8
species of  massive 2-d Majorana fermions
on a  2-d surface with a curved metric equal to the induced metric
with a ``$\sigma_3$'' mass term.
 The
square of the  resulting fermionic operator
$ \Delta_F=  - \hat \nabla^2 + \four R^{(2)}  + 1 $
is the same as for  a  2-d fermion with  unit mass (cf. \stat).
Furthermore,  since  the induced
2-d metric is conformally flat,
  and the total conformal anomaly  must cancel out \dgt,
  one  is able to    rescale the fermions by
  $(-g)^{1/4} = \r'^{1/2}$,  transforming their  kinetic
  term into a flat-space  Dirac form. We then  get
  4+4  flat-space  2-d Majorana fermions
   with $\s$-dependent  masses   $m_F = \pm \r'$  (cf. \joy).

\bigskip

Let us now include the $\n$-dependence.
First, in the case of no rotation in \ads, i.e. the background \roo\
($\r=0,\ \k=\n$)
 we get in \fer,\hop:   $ \vr_0= \n (\G_0 + \G_9)$,
 $\vr_1 =0$.
It is then  natural to impose the kappa-symmetry
gauge  $\G^+ \theta^I=0, \ \G^\pm = \mp \G_0 + \G_9$.
We  finish  with the following quadratic term
in the  fermionic part of the action
\eqn\lcc{
L_F =  i \n  \left( \bar \theta^1 \G^- \del_+ \theta^1       +
\bar \theta^2 \G^- \del_- \theta^2  -
  2 \n  \bar \theta^1 \G^- \Pi   \theta^2 \right)  \ , }
  $$
    \Pi\equiv  i \G_* \G_0  = \G_{1234} \  , \ \ \ \ \ \
   \Pi^2 =1 \  .   $$
This  is  exactly     the same  as the full fermionic \lc GS action
\refs{\mett,\mets} in the \pw background of \blau,
which describes  4+4 2-d Majorana  fermions  with mass $m_F = \pm \n$.

When both $\n$ and $\w$ are non-vanishing,
we may again
transform $\r_\a$ into $\G_\a$  by  applying the  two $SO(1,2)$
rotations: a boost in (02) plane \soo, and  a similar boost in the
(09) plane (cf. \poo)
\eqn\ghpo{
S=  \exp (\ha \a_1  \G_0 \G_2 )\   \exp (\ha \a_2  \G_0 \G_9 )  \ ,   }
$$
\cosh \a_1\ = {\k \cosh \r
 \ov \sqrt {\r'^2 + \n^2} } \ , \ \  \  \ \ \ \ \ \
  \sinh \a_2\ = { \n\ov \r' }  \ .    $$
 Here $\r'$ is given by \rgo, so that
 $\a_1$ is as  in \poo\  and
 $d\a_1\ov d\s $ =$
 {  \k \w \r' \ov {\r'^2 + \n^2}   } $,
  $d\a_2\ov d\s $ =$
 -  { \nu  \r'' \ov  \r' \sqrt {\r'^2 + \n^2} } $.
 After the first boost
 $\vr_{\hat 0} = \sqrt{ 1 + {\n^2\ov \r'^2} }  \G_0 + { \n\ov \r' }
 \G_9$,  while  $\G_*$  in \form\
 unchanged;  after the second boost
  $\vr_{\a} = (\G_0, \G_1)$   while
  $$ \G_* \to \td \G_* = S^{-1} \G_* S =
  i ( \cosh \a_2   \G_0  -  \sinh \a_2
 \G_9) \G_{1234} \ . $$
Repeating the same steps as in the discussion of the $\n=0$ case
above, i.e. choosing  the $\Psi^1=\Psi^2$ gauge
($\Psi^I= S^{-1} \theta^I$)
and Weyl-rescaling the fermions ($\Psi \to \r'^{-1/2}\Psi$ )
  to put the  kinetic term
in the flat-space form  we get (cf. \frmk,\fuu,\you)
  \eqn\fuo{
L_F = 2i
 \left( \bar \Psi \tau^a \del_a \Psi   +
  i \r' \bar \Psi  \M  \Psi \right)   \,,}
\eqn\youi{  \M \equiv
 \ha \ep^{\a\b} \vr_\a \td \G_* \vr_\b
= \cosh\a_2  \ M =
i\r'^{-1}  \sqrt{ \r'^2 +     \n^2 } \   \G_{234}  \ . }
This action\foot{Note that  after the rescaling of the fermions
this
 action  corresponding to
 the  $\theta^1=\theta^2$ gauge has a regular limit $\r' \to 0$
that is equivalent to the  $\G^+ \theta^I=0$ action \lcc.}
 describes again a system of 4+4  2-d Majorana fermions
 with masses
 \eqn\mase{
 m_F = \pm \sqrt{ \r'^2 + \n^2 }  \ ,  }
 incorporating consistently   the two  special  cases
 $\n=0$ and $\r'=0$  discussed above.
 The  contribution  of fermions
 to the 2-d logarithmic divergences
thus  cancels out  the one of the conformal-gauge
bosonic fluctuations  in \hyhh,
as required by the conformal invariance
of the \adss string sigma model
\refs{\mt,\dgt}.

%%%%%%%%%%%%%%%%%%%%%%%%%%%%%%%%%%%%%%%%%%%%
\newsec{
 Quantum  correction to  energy spectrum  of rotating  string
  }
%%%%%%%%%%%%%%%%%%%%%%%%%%%%%%

Our aim is  to  compute the leading (1-loop)  corrections
to the energy of string excitations
above the rotating string  ground state
\sll.
As usual in semiclassical soliton quantization,
we shall consider a sector of string states
for which
  the expectation
values for the spin  $\hat S$ and the $\vp$-momentum
$\hat J$ operators  are  fixed  and
equal to  their  classical values
$  S_0$ and $J_0$  in \spi,\jje, i.e. as in \avv\ we assume
\eqn\jotb{
\langle\Psi| \hat J | \Psi\rangle  = J = J_0 \equiv
 \sql \nu \ , \ \ \ \
\ \ \
\langle\Psi| \hat S | \Psi\rangle  = S = S_0 \equiv  \sql \ss \ . }
 The expectation value for the ground state
  energy  is   shifted
from the classical value $E_0$ in \eee\  by the expectation value
of the 2-d  Hamiltonian for quadratic fluctuations (see (A.8))
\eqn\eww{
E_{\rm vac} = E_0  +  {1 \ov \k} \Delta E_{\rm 2d} \ ,\ \ \
 \ \ \ \ \ \
\Delta E_{\rm 2d} = \langle 0| \hat H_{\rm 2d}|0\rangle \ . }
In contrast to what happened for the BPS state
with $E=J$\  ($N_n|0\rangle=N_n |0\rangle_0 + O({ 1 \ov \sql})
= O({ 1 \ov \sql}) $ in  \expa)
here the 1-loop quantum  correction $\Delta E$  will
no longer vanish. We shall estimate
$\DE$ for large spin $\ss$  and $J=0$
  to the leading order in
$ { 1 \ov \sql}$    and will show that,
like the classical contribution \Elongi,
 it scales  as
 $\ln S$,
i.e. there  are no  stronger $\ln^k S$ corrections at the first
quantum  order.
This provides  support to the
conjecture of \gkp\ that the $\ln S$ behaviour seen
at the string level can be interpolated  to weak coupling,
i.e. that for large $\ss = { S\ov \sql}$  and $J=0$
one has
\eqn\ipp{
E- S \ap  f(\l) \ln { S\ov \sql}   \ , \ }
where  in the string  sigma model loop  expansion
(the value of $a_1$ will be computed below)
\eqn\fff{
f(\l)_{\l \gg 1}  = a_0 \sql + a_1  + a_2 { 1\ov \sql } + ... \ , }
\eqn\uuu{ a_0 = { 1 \ov \pi} \ , \ \ \ \ \  \ \ \ \
 a_1 \ap  - { 3 \ln 2 \ov  \pi}    \ , }
while  $f(\l)_{\l \ll 1} = b_1 \l + b_2 \l^2 + ...$
should follow from the  SYM perturbation theory
for the anomalous dimension of the corresponding operator
(see \gkp\ and refs. there).
It is crucial for consistency of the proposal of  \gkp\ that
both string-theory and gauge-theory perturbative
expansions  should not  contain  terms with higher  than first
powers of $\ln { S\ov \sql} $  (which  would   dominate over
the  leading-order $\ln  { S\ov \sql}$  result).

As in the   pure-boost case discussed in section 4,
starting with  ground state    one can then  build up a tower of
string oscillator states on top of it.
All of them will have the same $S$ and $J$ but different
oscillator occupation numbers and thus  different
quantum   energies.
There will be several sub-sectors  of
  states  created by
oscillators of different fluctuation fields in  the
quadratic action.  Considering
all possible cases    will produce a generalization
of the quantum spectrum \expa\ of small string oscillations
near the  boosted  $E=J$ state to the case of a
 non-zero spin.

In   section 5 we have determined the Lagrangian
for small fluctuations near the solution \sll:
it is the sum  of  the bosonic part  given
in the conformal gauge
by \onn\ (or \hipo) and \obb, and the fermionic part given by
\fuo.  It describes  a collection of fields  with
$\s$-dependent  mass terms. One should
diagonalise the corresponding  second-order
differential operators defined on the  cylinder $(\tau,\s)$,
quantise the ``oscillator'' modes
and  impose  the constraints \conn,\coonn,
determining the expression
for the energy  and the  analog of the  level matching condition.
While this was  straightforward to do in the
case  of the  $S=0, J\not=0$
solution  discussed in  section 4 where the bosonic \bos\ and
fermionic \lcc\   parts of the action had constant  coefficients,
this is   a complicated  task  in the \ads rotating case,
unless one  makes  certain  approximations.
Below we shall consider
 the same ``short string'' and ``long string'' limiting cases \gkp\
 of the rotating solution  as in  sections  3.1 and 3.2.
We shall   also  note  that since
all the fields in  the quadratic fluctuation
action \onn,\obb,\fuo\ have similar  masses,
  to capture  some  qualitative
features of the string oscillator  spectrum  it is
enough to consider the  contributions
 to the energy  coming from excitations
 of  one scalar field $\bar \b_i$ in \obb\
(describing  fluctuations in \ads 3-sphere
directions
 transverse to the rotation plane).

%%%%%%%%%%%%%%%%%%%%%%%%%%%%%%%%%%%%%%%%%%%
\subsec{\bf Ground state energy  shift}
%%%%%%%%%%%%%%%%%%%%%%%%%%%%%%

Let us start with estimating
the energy of the rotating solution in
the   ``long string''  limit.
For simplicity, we shall  set  $J$ (i.e.
the boost parameter $\nu$)  to zero and
to avoid
 dealing with ``longitudinal'' modes consider
   the bosonic fluctuation action  as  it
follows from the Nambu action in the static gauge \stat.
We shall  use that   in the ``long string''  limit
  ($\eta \ll 1$, $S\gg 1$, see \spp)
$\r'$ is  approximately constant  and
very large  ($\r''$ in \secc\ is small
since  $\k \approx \w$, see \kaplo).
In this case the induced 2-d curvature  term
$R^{(2)}$ in \cuu\  is  approximately zero
everywhere except a very small interval of $\s$ near the turning points
$\s=\pi/2$ and $\s= 3\pi/2$  where $\r'$  rapidly approaches
 zero.
As we have discussed  in section 5,
to  have  manifest  conformal invariance  (and to maintain
  equivalence  with the Polyakov string quantization)
 one should  actually
  omit  this  problematic $R^{(2)}$ term in \jou\ altogether.
Then  we finish  with  the following system  of 2-d fields
defined on  a flat 2-d  cylinder
and having  approximately constant masses
(see \stit,\fuo):
1 scalar with mass squared  $m^2_\p =  4 \r'^2$,
2 scalars with  $m^2_\b= 2 \r'^2$,
5 massless scalars, 8   Majorana fermions  with $m^2_F= \r'^2$.
Using \eww\foot{Note that in the static gauge \sta\ ($t= \k \tau$)
the space-time and the 2-d energy  are equal, up to
$\k$-factor.}
we then  finish with the following
estimate  for the leading
quantum correction to the  classical energy  \eee:
\foot{Though the masses do change
 near the turning points,  one is able  to argue
  that since the change
 occurs only during a very short interval   of $\s$,
 this does not,
 for large $\ss$,  significantly influence  the spectrum of the
 Laplace operators.
 Indeed, in  the long string limit the
  $\s$-dependent  part of the
 mass $m^2 \sim  \r'^2 $   can be
 approximated by the  sum of two delta-functions
 located at the turning points.
  The corresponding Schrodinger equation  can be solved
  explicitly,   and  one finds that the leading asymptotics
  at large $\k$  can be estimated  by ignoring
  $\s$-dependent terms in the masses.
  For constant masses
the expression for the
  quantum correction to
 the energy  becomes essentially the same as
  in section 6.5 in \dgt.
  Here we consider the theory on a flat
 cylinder $(\tau,\s)$, so  the  individual field
 contributions to the 2-d vacuum energy are  the same
 as in the theory  with the  kinetic term $-\del^2 + m^2$.
 This is an analog  of the 2-d vacuum energy
$ (D-2) \sum_{n=1}^\infty n =  (D-2) \zeta(-1)
= - {D-2\ov 12} $ in the closed bosonic string case:
it vanishes in the standard flat-space   GS string case
 when all masses are  set to zero.}
  \eqn\vacc{ \Delta E  \approx   { 1 \ov  \k }
	\sum_{n=1}^\infty
\big[ \sqrt{n^2 + 4 \k^2  } + 2 \sqrt{n^2 + 2 \k^2}
+ 5 \sqrt{n^2} - 8 \sqrt{n^2 + \k^2} \ \big]  +  O( {1 \ov \sql})
 \ ,    }
where we dropped the $n=0$ contribution 
in the 2-d vacuum energy 
$ \ha \sum_{n=-\infty }^\infty  | \omega_n|$ 
which is subleading at
large $\kappa$
and  used that  (see \rgo,\mos,\kaplo)
\eqn\whe{
 \r' \ap \k \ , \ \ \ \ \ \ \ \ \k
 \ap \ { 1 \ov \pi } \ln { \ss } \  \gg 1 \ .  }
In contrast to the  $S^5$-boost case
(or the  ``plane-wave'' spectrum \expa) where
 the 2-d  vacuum energies cancelled  out
between 2-d massive bosons and fermions, here
the effective 2-d supersymmetry is spontaneously broken
by the classical solution ($\r'\not=0$)
 so  that the vacuum energy is
finite  (due to mass-squared sum rule)
but  is non-zero.

To obtain  the leading term in $\Delta E$ at large
$\k$  we   observe that  it
 can be found by replacing the convergent  sum
 in \vacc\  by the  integral
$
\sum_{n=1}^\infty f(n) \ap \int_1^{\infty}\, dx\, f(x)$,
where
$
f(x)=\sqrt{x^2 + 4 \k^2 } + 2 \sqrt{x^2 + 2 \k^2}
+ 5 \sqrt{x^2} - 8 \sqrt{x^2 + \k^2}.
$
Computing the integral, we find  that its asymptotics
at large $\k$ is given by
\eqn\resu{
\int_1^{\infty}\, dx\, f(x) \ \approx \
-{3\ln 2}\ \k^2 + O(\k) \ .
}
Taking into account the factor of  $1\ov \k$
in \vacc, we finish with
\eqn\onelo{
\DE  \ap  -{3\ln 2\ov \pi} \ln \ss\ +   O({1 \ov \sql}) \ , }
i.e.  with the result for  the coefficient $a_1$  quoted in \uuu.
The most important conclusion  is the absence of
higher  powers of $\ln \ss$ at the
1-loop  order.

%NN
Let us now argue that the same should be true also at
 higher orders of string  inverse tension ($1\ov \sqrt \l$)
  expansion.
 Again, the space-time energy shift will be related to
 the 2-d energy by the $1\ov \k$ factor (see \eww), so we need
 to  show  that  $\Delta E_{\rm 2d}$ scales at most as
 $\k^2$ for large $\k \sim \ln \ss$.
 The  2-d quantum field theory in question contains a collection
 of massive fields  with   approximately constant
 (for large $\ss$, i.e. in the long string case) masses and
 proportional to $\k$ with interaction  terms containing
 two derivatives  or two powers of $\k$ (cf. \gyg).
 It may  be modelled by a theory
 $L= \sqrt \l [ (1 + n_1 \phi^2 + ...)  (\del \phi)^2  +
 m^2 ( \phi^2 + k_1 \phi^4 + ...) ] \ , \ \ \  m\sim  \k
 $ supplemented by
 fermionic terms that should cancel all UV divergences.
  The theory is defined on the cylinder $0 < \tau < T, \
   0 \leq \sigma < 2\pi L $,   where  $T\to \infty$
   and $L$ is fixed (above we had $L=1$).
  Since this theory must be UV finite,  dimensional
  considerations  imply that
  the  $n$-loop contribution to the
  corresponding 2-d effective  action
  vacuum energy  should  scale as
  $\G_l= T (\Delta E_{\rm 2d})_l  = ({1 \ov \sqrt \l})^{l-1}
  V m^2  f_l ( mL)$,
  where  $V= T L$ is the 2-d volume
   and $f_l$ is a finite function of dimensionless ratio
  of the two IR scales.\foot{In  our  case of several fields
  with different masses   $f$  will  also
  contain  finite ratios  of the  masses, but these will
   stay constant  in the limit of large $\k$.}
Since in  the infinite volume limit $L\to \infty$ the
function  $f_l$ should approach a finite constant $c_l$,
the same should be true also for fixed $L$ but for large $m$, i.e.
$f_l(mL)_{m \gg 1} \to  c_l$. As a result,  for large $m \sim \k$
 we get  $(\Delta E_{\rm 2d})_l \approx  b_l \k^2 $.
This implies that string corrections to the energy
of the rotating string solution
should produce a  non-trivial  function $f(\l) $
in \ink\   but should keep  $h(\l)=0$.
Combined  with the  expected absence
of the higher $\ln^k S$ terms on the gauge theory side
(see \gkp\ and refs. there)
this provides  a non-trivial check of the gauge
theory -- string theory correspondence.

%%%%%%%%%%%%%%%%%%%%%%%%%%%%%%%%%%%%%%%%%%%
\subsec{\bf Massive scalar contribution to the quantum spectrum}
%%%%%%%%%%%%%%%%%%%%%%%%%%%%%%
To find the spectrum  of excited  string
states  near  the  rotating and boosted  string ground
state  on  should  study   different
sectors of states   created   by oscillators
corresponding to different fluctuation fields
in the quadratic action. While in pure-boost case
all the ``transverse''  fluctuations were
decoupled and had  the same constant mass $\nu$
the situation  for $\ss \not=0$ is, in general,  more involved.
However, since
 all the fields in  the quadratic fluctuation
action \onn,\obb,\fuo\ have similar  masses, we
may  get a qualitative picture of the
spectrum by looking at few simplest  cases.
One is  a sub-sector of states  carrying excitations
of the  scalar field $\bar \b_i$ in \obb\
with the mass  given in  \joy, i.e.
\eqn\mbeta{
m^2_{_\b} = 2\r'^2 + \n^2
 =2(\k^2 - \w^2)\sinh^2\r  +    2\k^2 - \nu^2
\ .}
We will  estimate the contribution
of this field   to the ``excited''  part of the energy spectrum
in the two  special limits in the parameter space of
the classical
solution.
As explained in Appendix A, the correction  to the
classical energy
is essentially determined by the 2-d Hamiltonian
associated with the fluctuation Lagrangian.

\medskip
%%%%%%%%%%%%%%%%%%%%%%%%%%%%%%%%%%%%%%%%%%
{\bf Short string case}
%%%%%%%%%%%%%%%%%%%%%%%%%%%%%%%%%%%
\medskip

\noindent
Here $\oo^2-\k^2\approx 1$, $\eta \gg 1$
  and an  approximate solution  for
$\r(\s)$ is given by \shortsol.
Using \kapsh\  and \shortsol\  we can rewrite
  \mbeta\ as
\eqn\mbetai{
m^2_{_\b} \approx  -2(\k^2 - \nu^2)\sin^2\s  +
 2\k^2 - \nu^2 \ap
\nu^2+ {2\ov\eta}\cos^2\s\ .}
To find the spectrum  of  scalar modes
we expand $\bar \beta(\tau,\s)
= \sum_n b_n (\tau) \Phi_n (\s) $ where
\eqn\eigeni{
- \Phi_n''(\s)+ m^2_{_\b}(\s) \Phi_n(\s) = l^2_n \Phi_n(\s)\ ,
\ \ \ \ \  \ \ \    \Phi_n (\s) = \Phi_n (\s+ 2 \pi) \ . }
For  large $\eta$
the mass \mbetai\ is approximately constant
and we are back to the  case of the  standard
 massive oscillator  spectrum.  Computing the
first correction in   perturbation theory in
$1/\eta$,  we get ($ \langle \cos^2 \s \rangle = \ha$)
\eqn\speci{
l^2_n \approx  n^2 + \nu^2 + {1\ov\eta}\approx n^2 + \k^2  \ .}
Using the proportionality
relation between the space-time and the 2-d energy
(cf.  \eww,(A.8))  we find that the scalar $\bar \b$
 contribution
to the  energy of an excited
string state
is given by (cf. \expa)
\eqn\Eshorti{
E \approx  E_{\rm vac}  + {1\ov\kappa}
\sum_{n=-\infty}^{\infty}
\sqrt{n^2 + \k^2}\ N_n + O({ 1 \ov \sql})
\ ,}
where $\Ev$ is the ``ground state'' energy
($\ap E_0$ in \eee\
in the short string limit), and
$N_n$ are the occupation numbers for the oscillator
states  of $\bar \b$.

If   $\n$ is small   as in \hort,
${\nu^2} \ll \ss\ll 1 $, i.e. $\k^2 \ap 2 \ss$
(see \kapsh,\etash)
 we  find  to  the leading order ($n\ne
0$)
\eqn\typ{
E\approx \sql (
\sqrt{2\ss}+
{\nu^2\ov 2\sqrt{2 \ss}} )
+{1\ov \sqrt{2\ss}}
\sum_{n=-\infty}^{\infty}\ ( 1-{\nu^2\ov 4\ss}
+  {\ss\ov n^2} ) \ |n| N_n
+ ... \ ,  }
where we replaced $ \Ev$ by $E_0$ in \hort.
Since in the short string case $\ss \ll  1$
(see \etash),  the excited states can be very heavy.

In the opposite case of large  $\nu \gg 1$
when  $\k^2 \ap \n^2  + { 2 \ss\ov \n}$,
 the energy  \Eshorti\ of states with $ n \ll \n$
becomes (cf.  \goo,\mall)
\eqn\snn{
E\ap  \sql (  \n  + \ss  +{\ss\ov
2\nu^2})   +
\sum_{n=-\infty}^{\infty}\ ( 1  + { n^2  \ov 2 \n^2}
- { n^2 \ss \ov \n^5} ) \  N_n  + ... \ .   }
The first two terms in the second bracket  are the same
as in the spectrum
 \mall\  corresponding to  the case of  $\ss=0$.

\medskip
%%%%%%%%%%%%%%%%%%%%%%%%%%%%%%%%%%%%%%%%%%
{\bf Long string case}
%%%%%%%%%%%%%%%%%%%%%%%%%%%%%%%%%%%%%
\medskip

\noindent
To approximate the expression for the mass \mbeta\
in the limit of large $\r_0$ or small $\eta$  let us first
 describe the form of the solution for $\r(\s)$ in this
  case.  Introducing  the function
\eqn\oop{ y (\s)  = \eta\ \sinh^2\rho(\s)  \ ,  }
we observe that it  takes values from 0 to 1,
 and satisfies, according to  \rgo,\too\
\eqn\ylong{
{1\ov 2\sqrt{\k^2-\nu^2}}\int_0^{y(\s)} {dx\ov \sqrt{x(x+\eta)(1-x)}} =
\s\ ,\ \ \ \ \ \ \ \
\ y({\pi\ov 2})= \eta\sinh^2\rho_0 = 1\ .}
Computing the integral and solving the resulting equation at
small values
of
$\eta$, we get  $y(\s)$  for $0 \leq  \s <  {\pi \ov 2}$
 that
satisfies
$y(0)=0,  \ \ y({\pi\ov 2})=1 + O(\eta^2)$:
\eqn\longsol{
y(\s)\approx {1\ov 4(1+f)^2}\left[-\eta (1+f)
+ 8f+4f\sqrt{4+\eta +\eta
f}\right]\
,\ \ \ \ \ \    f(\s)
\equiv \left({\eta\ov 16}\right)^{1-{2\s\ov\pi}}\ .}
In  the limit $\eta\to 0$ the  solution $y(\s)$
goes to zero  as $\eta^{{\pi -2\s\ov\pi}}$
at  all  points  in $0 \leq  \s <  {\pi \ov 2}$
and is equal to 1 at $\s = {\pi\ov 2}$. This also means that
$y(\s )\ln^2\eta $ goes to 0  at all points $0\le \s < {\pi\ov 2}$.

Substituting the solution \longsol\ into \mbeta, and
taking into account \too,\kaplo, we get
\eqn\mbetai{
m^2_{_\b} \approx  \nu^2 + {2\ov\pi^2}(1 - y(\s))\ln^2\eta\
\approx
\nu^2+ {2\ov\pi^2}\ln^2\eta \ .}
Thus, we see that the fluctuations
$\bar \b_i$ around the long string have $\s$-independent and very heavy mass.
The energy of the excited states is  then given by (cf. \vacc)
\eqn\tii{
E \ap  E_{\rm vac}  + {1\ov\kappa}
\sum_{n=-\infty}^{\infty}\sqrt{n^2 + \nu^2 + {2\ov\pi^2}\ln^2\eta}
\ N_n \ .}
In the case of small $\n$, i.e. (see \spp) \
$\nu^2 \ll  \ln^2 \eta \approx \ln^2 { \ss} $ we get
(using \kaplo)
\eqn\yyt{
E \ap  E_{\rm vac}  + {\pi^2 \ov
2\sqrt{2}\ln^2\ss}
\sum_{n=-\infty}^{\infty} (n^2-\nu^2)  \ N_n \ . }
Note that we do not get
 $ O(\ln S)$ terms in the oscillator part of the spectrum.

In the opposite case when
$\nu^2 \gg \ln^2\eta \ap \ln^2{\ss\ov\nu}$
\eqn\hoii{
E \ap  E_{\rm vac}  + {1\ov 2\nu^2} \sum_{n=-\infty}^{\infty}
\ ( n^2
+ {1\ov \pi^2}\ln^2{\ss\ov\nu} )\ N_n  \ . }

%%%%%%%%%%%%%%%%%%%%%%%%%%%%%%%%%%%%%%%%%%%%%%
\bigskip
\noindent
{\bf Acknowledgements}
%%%%%%%%%%%%%%%%%%%%%%%%%%%%%%%

\noindent

We are grateful to M. Vasiliev  for a  useful discussion.
This work  was supported by the DOE grant
DE-FG02-91ER40690.
The work of A.~A.~T.  was also supported in part by
the  PPARC SPG 00613,
 INTAS  99-1590 and the Royal Society  Wolfson
 research merit award.

%%%%%%%%%%%%%%%%%%%%%%%%%%%%%%%%%%%%
\appendix{A}
{Expression for the energy in conformal gauge}
%%%%%%%%%%%%%%%%%%%%%%%%%%%%%%
To compute the 1-loop  correction to the
space-time energy of a given classical
string configuration we need
the  expression for the energy  in terms of the
fluctuation
fields, governed  by the quadratic fluctuation
action  like \onn,\obb.
As described on the example of the boosted solution
in section 4, one  is to solve the conformal gauge constraints,
expressing the energy in terms  of the  quantum
values of  space-time  momenta  or conserved charges
and a term quadratic in the fluctuation fields
(which turned out to be the 2-d  Hamiltonian  of the
transverse fluctuation fields,  see \jjj--\expa).
Quantizing the  fluctuation fields using the quadratic part
 of the string action in the conformal gauge,
one  may then be able  to compute the expectation value of the
energy  in a given quantum state.

Here we will derive the expression for the energy
for the case of the expansion near the   rotating
solution \sll.
In general, the space-time energy is the
zero mode of the momentum conjugated to the time variable $t$
\eqn\ener{
E = \sqrt{\l}\int {d\s\ov 2\pi}\, \ G_{tt}\  \pa_0 t\ ,
}
where  $G_{tt}\equiv - G_{00}$, i.e. in the
present \ads case \add\
$G_{tt}=\cosh^2\r$.  As usual, the  l.h.s. of the
first  of the  two conformal gauge
constraints \conn\  is proportional to
 the density of the full 2-d  Hamiltonian
 $$
\H (t, \phi, \vp,\r, ... ) =
{1\ov 2} G_{MN} (X) \left( \pa_0 X^M  \pa_0 X^N  + \pa_1
X^M
\pa_1   X^N \right) $$
\eqn\hamil{
 = -{1\ov 2}G_{tt}\left( \pa_0 t \pa_0 t + \pa_1 t \pa_1
t\right) + ...
\ .}
Splitting  the  field
$t$ as $\k \ta + \td{t}$, and taking
into
account that the metric is invariant under the time translations, we get
from the constraint $\H =0$
\eqn\dott{
G_{tt} \pa_0 t = {1\ov 2} \k G_{tt} + {1\ov \k} \H (\td{t}, \phi,\vp,\r,
 ...
)
\ ,}
where $\H (\td{t}, \phi, \vp,\r, ... )$ is given by \hamil\ with the
replacement
$t\to \td{t}$. In addition to the energy,
our background has two other
conserved charges related to  the invariance of the metric
under translations along $\phi$ and $\vp$ directions
(cf. \spi,\jje):
\eqn\Smom{
S = \sqrt{\l}\int {d\s\ov 2\pi}\, G_{\p\p}\ \pa_0 \p\ ,
\ \ \ \ \ \
J = \sqrt{\l}\int {d\s\ov 2\pi}\, G_{\vp\vp}\ \pa_0 \vp\ ,
}
where $G_{\p\p}=\sinh^2\r\ \cos^2\b_1 \cos^2\b_2$ and
$G_{\vp\vp} = \cos^2\psi_1\cos^2\psi_2\cos^2\psi_3\cos^2\psi_4$.
Since the conformal-gauge
string  action and thus  the Hamiltonian
 depend quadratically on $\p$ and $\vp$, we may repeat the
 same splitting procedure  with $\p$ and $\vp$  in
 \dott: setting
$\p = \oo \ta + \td{\p}, \  \vp = \nu \ta + \td{\vp}$,
using  the definitions \Smom\
and integrating \dott\  we arrive at the following expression
for the energy \ener:
\eqn\eneri{
E = {\oo\ov \k}S + {\nu\ov \k}J +
{\sqrt{\l}\ov \k}\int {d\s\ov 2\pi}\, \left[ {1\ov 2} \k^2G_{tt} - {1\ov
2}
\oo^2 G_{\p\p} - {1\ov 2} \nu^2 G_{\vp\vp} +
\H (\td{t}, \td{\phi}, \td{\varphi},\r ,\ldots ) \right]
\ , }
where $\H (\td{t}, \td{\phi}, \td{\varphi},\r ,\ldots )$ is again given
by the l.h.s. of
\hamil\ with the replacement $t\to \td{t}, \ \phi\to \td{\phi},\ \vp\to
\td{\varphi} $.

Next, let us expand  the remaining  fields  near their  classical
values as in \exx: $\r (\ta ,\s )= \r (\s )+ \kk
\td{\r}, \ \b_i =\kk  \td \b_i, \
\psi_s = \kk  \td \psi_s$ ($i=1,2; \ s=1,2,3,4$),
rescaling also $\td t,\td\p, \td \vp$ by $\kk$.
 To quadratic order in the fluctuations we then get
\eqn\energii{\eqalign{
E = &E_0+ {\oo\ov \k}(S-S_0) + {\nu\ov \k}(J-J_0)+
{1 \ov \k}\int {d\s\ov 2\pi}\,  \left[ {1\ov 2}
(\k^2-\oo^2)\cosh 2\r\  \td{\r}^2 \right. \cr
&
\left. +\ {1\ov 2}\oo^2\sinh^2\r \ \td \b_i^2 +
{1\ov 2}\nu^2 \td \psi_s^2
 + \H^{(2)} (\td{t}, \td{\phi}, \td{\varphi},
 \td{\r} ,\td \b_i, \td \psi_s )
\right] \ , }}
where $E_0,\ S_0,\ J_0$ are the ``classical''
values of  energy and charges computed on the rotating
string background, i.e.  given by \eee,\spi,\jje.
Comparing this to  the action \onn,\obb\ or, equivalently, \hipo,
it is easy to check that the integrand in
\energii\ coincides with the 2-d  Hamiltonian $  \H_{\rm 2d} $
corresponding to the quadratic  fluctuation action
  \onn,\obb\ or, equivalently, \hipo. The explicit
    expression for $  \H_{\rm 2d} $  in terms of fields and their
    derivatives
   can be obtained  from  the (minus) Lagrangian  by  dropping
   the  terms
linear in  time derivative
(i.e. mixed terms in \onn), and  by reversing the  signs of the
terms quadratic in  time derivative. In terms of redefined fields in
\onn,\obb\
$$
 \H_{\rm 2d} =
 \ha  \big[
-  \del_0 \bar t \del_0 \bar  t  -  \del_1 \bar t \del_1 \bar  t
  - \m^2_t \bar t^2
+   \del_0 \bar  \p  \del_0  \bar \p
+   \del_1 \bar  \p  \del_1 \bar \p + \m^2_\p \bar \p^2
 $$  $$
+ \   \del_0 \rr \del_0 \rr      +   \del_1 \rr \del_1 \rr
      + \m^2_\r  \rr ^2  +
 \del_0 \bar \b_i   \del_0 \bar \b_i   +
 \del_1\bar \b_i   \del_1 \bar \b_i
 + m^2_{_\b} \bar\b_i^2   $$ \eqn\expli{
+\     \del_0 \td \vp \del_0 \td \vp  + \del_1 \td \vp \del_1  \td \vp
+  \del_0 \td \psi_s \del_0  \td \psi_s   +
  \del_0 \td \psi_s \del_0  \td \psi_s +   \nu^2
    \td \psi_s^2  \big] \ .
}
 The final form of
  \energii\  is thus
\eqn\energiii{
E = E_0+ {\oo\ov \k}(S-S_0) + {\nu\ov \k}(J-J_0) +
{   1  \ov \k}\int {d\s\ov 2\pi}\,
\H_{\rm 2d} (\td{t}, \td{\phi}, \td{\varphi},\td{\r} ,
\td \b_i, \td \psi_s
 )  \ . }
This expression generalises  the one in the pure boost case,
which, upon quantisation led to  eq.\expa\
(there $\w=0, \ \k=\n, \ S=S_0=0, \ E_0=J_0$).
In general,
assuming one can  diagonalise
the second-order differential operators
appearing in \onn,\obb,
one can then choose a specific
quantum state (subject to
the ``level matching'' condition
following from the second constraint \coonn)
 with given  quantized values
of $J$ and $S$  equal to their classical values
(see \jotb)
and compute  the average value
of $E$ in \energiii:
\eqn\energiii{
E = \langle\Psi| \hat E | \Psi \rangle  = E_0+  { 1 \ov \k}
\langle\Psi| \hat  H_{\rm 2d}  | \Psi\rangle  \ , }
$$ H_{\rm 2d} \equiv
\int {d\s\ov 2\pi}\,
\H_{\rm 2d} (\td{t}, \td{\phi}, \td{\varphi},\td{\r} ,
\td \b_i, \td \psi_s
 )  \ . $$
Finally, one may
express   the classical parameters
$\n,\w,\k$ in terms of $S$ and $J$  in the
leading quantum  correction (i.e.  up to  $O({1 \ov \sql})$ terms
as in  \expa)  using the
classical relations  of section 3.

%%%%%%%%%%%%%%%%%%%%%%%%%%%%%%%%%%%%%%%%%%%%%%%%%%%%%%%%
\vfill\eject
\listrefs

\end